\documentclass[a4paper,onecolumn,11pt,accepted=2025-04-02]{quantumarticle}
\pdfoutput=1
\usepackage[utf8]{inputenc}
\usepackage[english]{babel}
\usepackage[T1]{fontenc}
\usepackage{amsmath}
\usepackage{hyperref}

\usepackage[numbers,sort&compress]{natbib}

\usepackage{tikz}
\usepackage{lipsum}

\usepackage[utf8]{inputenc}
\usepackage{amsmath}
\usepackage{multirow}
\usepackage{tikz}
\usepackage{hyperref}

\usepackage{wrapfig}

\usepackage{breqn}
\makeatletter
\newcommand*{\rom}[1]{\expandafter\@slowromancap\romannumeral #1@}
\makeatother

\usetikzlibrary{quantikz}

\usepackage{graphicx}
\usepackage{lipsum}
\usepackage[utf8]{inputenc}
\usepackage[english]{babel}

\usepackage{graphicx}
\usepackage{dcolumn}
\usepackage{bm}
\usepackage{adjustbox}

\usepackage{graphicx}
\usepackage{dcolumn}
\usepackage{bm}

\usepackage{graphicx}
\graphicspath{{C:..}}
\usepackage{dcolumn}
\usepackage{bm}

\usepackage{graphicx}
\usepackage{amsmath}
\usepackage{amssymb}
\usepackage{bm}
\usepackage{epsfig}
\usepackage{dcolumn}
\usepackage{slashed}
\usepackage{empheq}

\newcommand{\bea}{\begin{eqnarray}}
\newcommand{\eea}{\end{eqnarray}}

\begin{document}

\title{$\mathcal{PT}$-symmetric mapping of three states and its implementation on a cloud quantum processor}

\author{Yaroslav Balytskyi}
\affiliation{Department of Physics and Astronomy$,$ Wayne State University$,$ Detroit$,$ MI$,$ 48201$,$ USA}
\email{ybalytsk@uccs.edu}
\email{hr6998@wayne.edu}
\orcid{0000-0002-7582-6232}

\author{Yevgen Kotukh}
\affiliation{Department of Information Technologies$,$ Yevhenii Bereznyak Military Academy$,$ Kyiv$,$ 04050$,$ Ukraine}
\email{yevgenkotukh@gmail.com}
\orcid{0000-0003-4997-620X}

\author{Gennady Khalimov}
\affiliation{Department of Information Security$,$ Kharkiv National University of Radio Electronics$,$ Kharkiv$,$ 61166$,$ Ukraine}
\email{hennadii.khalimov@nure.ua}
\orcid{0000-0002-2054-9186}

\author{Sang-Yoon Chang}
\affiliation{Department of Computer Science$,$  
University of Colorado$,$ Colorado Springs$,$ Colorado 80918$,$ USA}
\email{schang2@uccs.edu}
\orcid{0000-0002-5736-5823}

\maketitle

\begin{abstract}
Recently, $\mathcal{PT}$-symmetric systems have garnered significant attention due to their unconventional properties. Despite the growing interest, there remains an ongoing debate about whether these systems can outperform their Hermitian counterparts in practical applications, and if so, by what metrics this performance should be measured. We developed a novel $\mathcal{PT}$-symmetric approach for mapping $N = 3$ pure qubit states to address this, implemented it using the dilation method, and demonstrated it on a superconducting quantum processor from the IBM Quantum Experience. For the first time, we derived exact expressions for the population of the post-selected $\mathcal{PT}$-symmetric subspace for both $N = 2$ and $N = 3$ states. When applied to the discrimination of $N = 2$ pure states, our algorithm provides an equivalent result to the conventional unambiguous quantum state discrimination. For $N = 3$ states, our approach introduces novel capabilities not available in traditional Hermitian systems, enabling the transformation of an arbitrary set of three pure quantum states into another, at the cost of introducing an inconclusive outcome. Our algorithm has the same error rate for the attack on the three-state QKD protocol as the conventional minimum error, maximum confidence, and maximum mutual information strategies. For post-selected quantum metrology, our results provide precise conditions where $\mathcal{PT}$-symmetric quantum sensors outperform their Hermitian counterparts in terms of information-cost rate. Combined with punctuated unstructured quantum database search, our method significantly reduces the qubit readout requirements at the cost of adding an ancilla, while maintaining the same average number of oracle calls as the original punctuated Grover's algorithm. This provides significant advantages for NISQ-era computers. Our work opens new pathways for applying $\mathcal{PT}$ symmetry in quantum communications, computing, and cryptography.
\end{abstract}

\bigskip
\noindent\textbf{Keywords:} $\mathcal{PT}$-symmetric transformations; Quantum state discrimination; Technical advantages; Quantum sensing; Information-cost rate; Quantum database search; Qubit readout cost; IBM Quantum Experience; Quantum key distribution.

\tableofcontents

\section{Introduction}

The problem of identifying information stored in a quantum system is fundamental in quantum computer science, and the simplest option is to use two-dimensional systems or qubits to store quantum information. In classical physics, the system's state variables are also observables and there is no fundamental limitation on the precision with which they can be determined. By contrast, the quantum observables are represented by operators acting on the vector space, which in the general case do not commute, and the outcome of the measurement has statistical rather than deterministic properties. Quantum systems in different non-orthogonal states are impossible to perfectly distinguish even when arbitrarily large but finite number of samples for the measurements are available~\cite{helstrom1969quantum,holevo2011probabilistic,bergou2010discrimination,chefles2000quantum,barnett2009quantum,bae2015quantum}. In other words, no test can guarantee a correct guess every time, a fact highlighted by the quantum Chernoff bound~\cite{audenaert2007discriminating}. 

Quantum state discrimination involves two parties who agree on a set of allowed states in which the system can be, and their prior probabilities of occurrence. A measurement can obtain only a finite amount of information, and thus this set must be finite. Alice prepares a state from this set and sends it to Bob, who must determine it using the appropriate measurement. Quantum state discrimination has several important applications. In particular, it is strongly linked to a dimension witness of quantum systems~\cite{brunner2013dimension} and represents an operational interpretation of conditional mutual entropy~\cite{konig2009operational}. The quantum key distribution (QKD) security is based on the hardness of quantum state discrimination and the no-cloning theorem~\cite{wootters1982single}. The search over an unstructured database can be mapped to the discrimination of the states exponentially close to each other~\cite{abrams1998nonlinear}. 

Quantum state discrimination is difficult apart from the $N = 2$ case, and the existing strategies for quantum state discrimination can be classified into the minimum error discrimination~\cite{helstrom1969quantum}, unambiguous discrimination~\cite{ivanovic1987differentiate,dieks1988overlap}, and maximum confidence discrimination~\cite{croke2006maximum}, each with its advantages and drawbacks. The minimum error discrimination solution was obtained for the states possessing particular symmetries such as ``geometrically uniform'' states~\cite{eldar2004optimal}, and mirror-symmetric states~\cite{andersson2002minimum}. In the general case of $N = 3$ states, the minimum error discrimination solution for pure qubit states was obtained in~\cite{hunter2004results,samsonov2009minimum}. In the general case, however, the solution requires intricate computation. Unambiguous state discrimination can be achieved only for linearly independent states~\cite{chefles1998unambiguous}, and therefore is impossible for $N = 3$ qubit states.

Meanwhile, recent developments in $\mathcal{PT}$-symmetric quantum mechanics~\cite{bender2002complex,bender2003must,mostafazadeh2002pseudo}, where the condition of Hermiticity is replaced by the condition of $\mathcal{PT}$-symmetry, provide new opportunities for the quantum information science, that are not available in the usual Hermitian case. Such theories possess an additional degree of freedom represented by the $\alpha$ parameter, and in the limit $\alpha \rightarrow 0$,  the intersection of $\mathcal{PT}$-symmetric and Hermitian cases are real symmetric Hamiltonians~\cite{bender2003finite}. At certain values of the $\alpha$ parameter, the degeneracies occur, known as exceptional points, which correspond to coalescing eigenvectors and eigenvalues~\cite{miri2019exceptional,ozdemir2019parity}. 

These points can be used in multiple applications. First, in the $\mathcal{PT}$-symmetric system, the time of quantum evolution may approach \textit{theoretically zero} near the exceptional point, while a finite time is needed in the Hermitian one. This effect was demonstrated both theoretically~\cite{bender2007faster} and experimentally~\cite{zheng2013observation}. Second, it was used for enhanced sensing~\cite{wiersig2014enhancing,chen2017exceptional,hodaei2017enhanced,langbein2018no,lau2018fundamental,zhang2019quantum}, and it have been shown that $\mathcal{PT}$-symmetric sensors are 8.856 times superior to Hermitian ones~\cite{yu2020experimental}. Moreover, $\mathcal{PT}$-symmetric operations increase the Quantum Fisher information (QFI) needed to increase the accuracy of quantum parameter estimation~\cite{guo2017enhancing,wang2020quantum}, which in turn was used for Bayesian parameter estimation~\cite{balytskyi2021detecting,balytskyi2021PTBayes,balytskyi2022}. 

Exceptional points have also been experimentally realized in superconducting transmon circuits, demonstrating their utility for quantum state control and non-Hermiticity-induced chiral state transfer~\cite{naghiloo2019,abbasi2022,chen2022}. Additionally, exceptional points have been observed in a variety of classical systems~\cite{dembowski2001experimental,ruter2010observation,peng2014parity,schindler2012symmetric,bender2013observation,shi2016accessing,zhu2014p,li2019observation,partanen2019exceptional}, where they have found applications in laser mode management~\cite{peng2016chiral,brandstetter2014reversing,wong2016lasing} and topological mode transfer~\cite{xu2016topological,doppler2016dynamically,choi2017extremely,zhang2018dynamically}. 

For $N = 2$ states, $\mathcal{PT}$-symmetric quantum state discrimination was developed in \cite{bender2013pt}. However, $\mathcal{PT}$-symmetric transformation inherently requires postselection, and no formula for the probability of successful postselection was provided in \cite{bender2013pt}. Our work extends this method to $N = 3$ states and, for the first time, derives exact expressions for the population of the postselected space for both $N = 2$ and $N = 3$.

The applicability of our results extends beyond quantum state discrimination, allowing us to identify scenarios and conditions in which $\mathcal{PT}$-symmetric systems outperform their Hermitian counterparts. As discussed further in the text, $\mathcal{PT}$-symmetric metrology can be viewed as part of a broader class of postselected metrology, which is capable of producing anomalously large information-cost rates~\cite{arvidsson2020quantum}. However, previous research~\cite{arvidsson2020quantum} did not explicitly identify the probability of successful postselection $\left(p^{ps}\right)$ and the effective QFI rescaled by this probability $\left(p^{ps}_{\theta}\mathcal{I}^{ps}\right)$, which together define the information-cost rate. Our results, detailed further in the text, enable precise quantification of the information-cost rate in $\mathcal{PT}$-symmetric postselection qubit metrology, providing experimentalists with the necessary tools to evaluate the performance of such systems.

Additionally, Grover's search algorithm operates effectively within a two-dimensional subspace~\cite{grover1996proceedings}, making our results directly applicable to this scenario. By integrating our $\mathcal{PT}$-symmetric approach with the punctuated version of Grover's algorithm~\cite{boyer1998tight,gingrich2000generalized}, we demonstrate a significant reduction in qubit readout costs at the expense of introduction of an ancilla, used only once to perform the $\mathcal{PT}$-symmetric transformation. This reduction is achieved while preserving the same average number of oracle calls as the original punctuated version of Grover's algorithm. Consequently, our method enhances the efficiency of quantum searches, particularly for NISQ-era computers~\cite{preskill2018quantum}, by optimizing resource usage without compromising performance.

\section{Results and structure of the paper}
\label{Results}

First, we provide the necessary background on $\mathcal{PT}$ symmetry in Section~\ref{Background}. In Section~\ref{ThreeStatesTheory}, we develop a $\mathcal{PT}$ symmetric approach for the  $N = 3$ pure quantum states discrimination, which consists of two stages of $\mathcal{PT}$ symmetric evolution. In the first one, two of the three states are made mutually orthogonal in terms of the Hermitian scalar product. The second stage enables the transformation of a given set of three arbitrary states into another set of states as required. In the limiting case as $\alpha$ approaches $\pm\frac{\pi}{2}$, near the exceptional point, the geometry of the postselected space closely resembles that of a two-state scenario. Our initial findings regarding the $\mathcal{PT}$-symmetric subsystem for $N = 3$ states~\cite{balytskyi2020mathcal} were validated through experiments conducted on an optical setup~\cite{chen2022quantum} (see Section III and Ref. [38] therein for further information). In Section~\ref{Embed}, we derive, for the first time, exact expressions for the population of the postselected $\mathcal{PT}$-symmetric subspace for both $N = 2$ and $N = 3$. In contrast, prior studies have relied on numerical computations and experimentation~\cite{chen2022quantum,wang2024demonstration,wu2019observation,dogra2021quantum} to embed the $\mathcal{PT}$-symmetric subsystem into a Hermitian Hamiltonian. We show that for $N = 2$ states, the proposed method has the same probability of success as conventional unambiguous quantum state discrimination~\cite{ivanovic1987differentiate}. This fact was experimentally determined in~\cite{chen2022quantum} and confirmed in a more recent experimental work~\cite{wang2024demonstration}, which also found that this procedure, while having the same success probability, requires fewer quantum resources compared to the regular Hermitian case. Our work provides the missing analytical derivation of this fact. The corresponding results for $N = 3$ are used further in the text to identify applications and relevant metrics where $\mathcal{PT}$-symmetric systems outperform their Hermitian counterparts.

For $N = 3$, in Section~\ref{Experimental}, we provide a comparison between our theoretical model and the results of its implementation on IBM Quantum Experience. The details of the implementation are provided in Methods~\ref{Methods}. In Section~\ref{TrineQKD}, we apply our algorithm to an attack on the trine state QKD protocol~\cite{phoenix2000three} and demonstrate that our algorithm achieves the same error rate as the minimum error, maximum confidence, and maximum mutual information strategies.

In Section~\ref{SensingApplication}, we apply our results to quantum sensing with non-Hermitian $\mathcal{PT}$-symmetric gates. Such a setup promises divergent susceptibility~\cite{chu2020quantum}, which is relevant for sensing applications. However, it has been shown that the QFI under postselection does not increase~\cite{combes2014quantum}, indicating that $\mathcal{PT}$-symmetric quantum sensing \textit{cannot} outperform Hermitian sensing in terms of QFI. Additionally, when rescaled by the probability of a successful outcome, the effective susceptibility does not diverge, as most measurements do not yield meaningful information about the parameter of interest~\cite{ding2023fundamental}. While the results in~\cite{ding2023fundamental} suggest that non-Hermitian sensors do not outperform their Hermitian counterparts when resources are unlimited, they also indicate that this conclusion may not hold in practice due to the inevitable limitation of resources. In Section~\ref{SensingApplication}, we explicitly compute the QFI rescaled by the maximal probability of successful postselection and show that, at the exceptional point, it is the same as in the Hermitian case, thus aligning with previous research.

At the same time, as demonstrated in~\cite{arvidsson2020quantum}, postselected quantum metrology can outperform Hermitian metrology in terms of the information-cost rate when the measurement cost is sufficiently high, yielding better results in resource-constrained scenarios. However, the precise expression for the QFI rescaled by the probability of successful postselection was not found in~\cite{arvidsson2020quantum}, but our work provides this crucial detail. These exact formulas enable the identification of specific conditions -- considering preparation, postselection, and measurement costs -- under which a $\mathcal{PT}$-symmetric sensor outperforms its Hermitian counterpart. This provides experimentalists with the necessary tools to evaluate and implement $\mathcal{PT}$-symmetric sensors, optimizing their performance and taking full advantage of the benefits offered by $\mathcal{PT}$-symmetry in various quantum metrology applications.

In Section~\ref{SearchApplication}, we consider our approach in the context of an unstructured quantum database search. Previous studies~\cite{bender2013pt,abrams1998nonlinear,croke2015pt} discussed the potential to improve unstructured quantum database searches by applying $\mathcal{PT}$ symmetry. Since the advantages provided by $\mathcal{PT}$ symmetry come at the price of introducing an inconclusive outcome, it is impossible to reduce the average number of oracle calls~\cite{bender2013pt,abrams1998nonlinear,croke2015pt} to find the marked element in the database, in agreement with our results. 

However, as we discuss in the same Section~\ref{SearchApplication}, in addition to the number of oracle calls, the \textit{circuit depth} and the \textit{qubit readout cost} are critical parameters that can limit the performance of near-term (NISQ-era) quantum hardware. We demonstrate that, when combined with the punctuated unstructured quantum database search, our algorithm significantly reduces the qubit readout demands while preserving the average number of oracle calls. By introducing a single ancilla qubit -- used only once at the conclusion of a depth-limited Grover's search -- our approach renders the depth–measurement trade-off discussed in~\cite{ng2024} \textit{unnecessary} since full measurement of all working qubits is performed \textit{only once}, when a single ancilla passes the postselection.

Our results highlight the practical advantages of $\mathcal{PT}$-symmetric approaches in quantum computing, offering means to optimize resource usage and improve performance in real-world applications. We present our conclusions and outline future work in Section~\ref{Conclusions}. For details on numerical input for processors from IBM Quantum Experience, see Methods~\ref{Methods}.

\section{Background on \texorpdfstring{$\mathcal{PT}$}{PT} symmetry}\label{Background}

For a complete description of the physical system, the energy eigenvalues of its Hamiltonian must be real-valued. Complex energies are often used to describe dissipative phenomena when the probability of finding a particle decreases over time. However, the decaying particle does not vanish but transforms into other particles, making this description incomplete. The condition of reality of the spectra can be achieved by constraining the Hamiltonian to be Hermitian, $H = H^\dagger$. However, this condition is not necessary and can be replaced by a condition of unbroken $\mathcal{PT}$-symmetry~\cite{bender2002complex,bender2003must,mostafazadeh2002pseudo}, which guarantees that all eigenvalues of the Hamiltonian are real. Additionally, it provides an extra degree of freedom not available in the Hermitian case, which we describe further.

The Hamiltonian is called $\mathcal{PT}$-symmetric if it satisfies the condition $\mathcal{H} = \mathcal{H}^{\mathcal{PT}}$. The signs of the quantum mechanical coordinate and momentum, $\hat{x}$ and $\hat{p}$, are changed by the parity operator $\mathcal{P}$ as $\mathcal{P}\hat{x}\mathcal{P} = - \hat{x}$,  $\mathcal{P}\hat{p}\mathcal{P} = - \hat{p}$, and for the case of qubit, up to a unitary transformation, it is given by~\cite{bender2007faster}:\begin{equation}\label{P}
\mathcal{P} = 
\begin{pmatrix}
0 & 1 \\
1 & 0
\end{pmatrix}   
\end{equation}
The time-reversal operator $\mathcal{T}$ changes the signs of the imaginary unit and the momentum operator as $\mathcal{T}i\mathcal{T} = - i$ and $\mathcal{T}\hat{p}\mathcal{T} = - \hat{p}$. The $\mathcal{PT}$ operator is a combination of $\mathcal{P}$ and $\mathcal{T}$. For the case of qubit, the most general $\mathcal{PT}$-symmetric Hamiltonian depends on three real parameters, $r$, $s$ and $\beta$ as~\cite{bender2013pt}: \begin{equation}
\mathcal{H} = \mathcal{H}^{\mathcal{PT}} = 
\begin{pmatrix}
r e^{i\beta} & s \\
s & r e^{-i\beta}
\end{pmatrix}\label{Hamiltonian1}   
\end{equation}

The $\mathcal{PT}$-symmetric Hamiltonian is called to possess an unbroken $\mathcal{PT}$ symmetry if each of its eigenfunctions is also an eigenfunction of the $\mathcal{PT}$ operator. This condition guarantees that all energy eigenvalues are real~\cite{dorey2001spectral, dorey2004reality}. Additionally, this condition provides an extra $\mathcal{C}$ operator that is not available in the Hermitian case. This operator is represented by the sum of all eigenfunctions of the $\mathcal{PT}$-symmetric Hamiltonian in Eqn.~\eqref{Hamiltonian1}:
\begin{equation}
\mathcal{H}\psi_n\left(x\right) = E_n\psi_n\left(x\right), \ \mathcal{C}\left(x, y\right) = \sum_{n=1}^{2} \psi_n\left(x\right)\psi_n\left(y\right)
\end{equation}
For the qubit case, it takes the form:
\begin{equation}\label{C_operator}
\mathcal{C} =\frac{1}{\cos\left(\alpha\right)}
\begin{pmatrix}
i \sin\left( \alpha \right) & 1 \\
1 & -i \sin\left( \alpha \right)
\end{pmatrix},    
\end{equation}
with $\alpha$ being expressed as $
\sin\left(\alpha\right) = \frac{r}{s}\sin\left(\beta\right)
$. As a result, the set of commuting operators in the $\mathcal{PT}$-symmetric theory is bigger compared to the Hermitian case, $\left[\mathcal{C}, \mathcal{H}\right] = 0$ and $\left[\mathcal{C}, \mathcal{PT} \right] = 0$. 

An arbitrary $\mathcal{PT}$-symmetric Hamiltonian $\mathcal{H}$ can be mapped into the Hermitian one $\mathcal{H}'_{\mathrm{Hermitian}}$ using the following mathematical transformation:
\begin{equation}
\label{MappingPTtoHerm}
\mathcal{H}'_{\mathrm{Hermitian}} = e^{-\frac{Q}{2}} \cdot \mathcal{H} \cdot e^{\frac{Q}{2}}, \quad \text{where } e^{Q} = \mathcal{C} \mathcal{P},
\end{equation}
since $\mathcal{C}\,\mathcal{P}$ is a positive and invertible operator~\cite{mostafazadeh2002pseudo,mostafazadeh2003jphysa}.
From the physical point of view, as discussed in~\cite{gardas2016repeatability}, a quantum system can be represented either by a Hermitian Hamiltonian or by a non-Hermitian Hamiltonian with a real spectrum. The transformation between these two pictures can be interpreted as a ``quantum Coriolis force'', analogous to switching to a non-inertial reference frame in classical physics~\cite{gardas2016repeatability}.

In general, $\mathcal{PT}$-symmetry is a partial case of the broader concept of pseudo-Hermiticity
~\cite{mostafazadeh2002pseudo,mostafazadeh2002pseudo1,mostafazadeh2005,bagchi2002pla, ahmed2002pla1, japaridze2002jphysa, ahmed2003ensemble, ahmed2003cpt, ahmed2003pre, ahmed2003jphysa, ahmed2003pseudo1, ahmed2003cpt1, blasi2004pseudo, bagchi2005pseudo}. For a detailed review, the interested reader should refer to~\cite{bender2007nonhermitian}.
A linear operator $O$ is called \textit{pseudo-Hermitian} if there exists a Hermitian \textit{intertwining operator} $\eta$ such that:
\begin{equation}
    O^\dagger \;=\; \eta \cdot O \cdot\eta
\end{equation}
The $\mathcal{PT}$-symmetric Hamiltonian in Eqn.~\eqref{Hamiltonian1} is a special case of pseudo-Hermiticity, since the parity operator $\mathcal{P}$ in Eqn.~\eqref{P} is Hermitian and thus can serve as the intertwining operator:
\begin{equation}
    \mathcal{H}^\dagger \;=\; \mathcal{P}\cdot\mathcal{H}\cdot\mathcal{P}.
\end{equation}
Moreover, the Jarzynski equality~\cite{jarzynski1997nonequilibrium} -- originally derived in Hermitian settings to relate nonequilibrium work and equilibrium free energy difference -- also extends to systems with unbroken $\mathcal{PT}$-symmetry~\cite{deffner2015jarzynski} and, even more generally, to non-Hermitian systems with real spectra~\cite{gardas2016nonhermitian}. It should be noted that the Carnot bound~\cite{carnot1824} holds even in systems with a spectrum containing complex eigenenergies if they consist of complex-conjugate pairs~\cite{gardas2016nonhermitian}.

Despite these parallels between them, the $\mathcal{PT}$-symmetric and Hermitian descriptions are \emph{mathematically distinct} because the mapping in Eqn.~\eqref{MappingPTtoHerm} is \emph{nonunitary}. As a result, orthogonal eigenvectors in one representation can become non-orthogonal in the other, leading to phenomena such as faster-than-Hermitian quantum evolution~\cite{bender2007faster}.

In Sections~\ref{SensingApplication} and~\ref{SearchApplication}, we further demonstrate that this difference is more than a purely mathematical subtlety. In particular, employing 
$\mathcal{PT}$-symmetry can yield quantifiable \textit{technical} advantages over the Hermitian case, indicating that these two frameworks are not fully equivalent in practical applications.

One particular implementation of a $\mathcal{PT}$-symmetric qubit was studied in~\cite{gardas2016ptdecoherence}, where a $\mathcal{PT}$-symmetric subsystem governed by a non-Hermitian Hamiltonian $\mathrm{H}_S \neq \mathrm{H}_S^\dagger$ was coupled to a Hermitian one by $\mathrm{H}_S\otimes\mathrm{V}_B$. Both $\mathrm{H}_B$ governing the Hermitian environment and $\mathrm{V}_B$ were assumed to be Hermitian. This setup was shown to be more robust against the decoherence~\cite{gardas2016ptdecoherence} than its Hermitian counterparts. The total non-Hermitian Hamiltonian:
\begin{equation}
    \mathrm{H}_{S+B}
    \;=\;
    \mathrm{H}_S\otimes\mathrm{I}_B
    \;+\;
    \mathrm{I}_S\otimes\mathrm{H}_B
    \;+\;
    \mathrm{H}_S\otimes\mathrm{V}_B,
\end{equation}
is transformed to the Hermitian representation using Eqn.~\eqref{MappingPTtoHerm}, and after obtaining the solution, the inverse map is applied to obtain the final solution.

In contrast, in Section~\ref{Embed}, we embed the $\mathcal{PT}$-symmetric subsystem in such a way that the \textit{total} system \underline{remains} Hermitian. Such an implementation is easier to realize experimentally~\cite{wu2019observation} and straightforward to build on a superconducting quantum processor~\cite{dogra2021quantum}. The $\mathcal{PT}$-symmetric transformation succeeds with a certain probability, which we maximize by choosing appropriate parameters, and thus functions as a single \textit{probabilistic quantum gate}. Although it has been argued that:``Probabilistic quantum gates are to be avoided wherever possible, as the probability of success of a circuit composed of such gates falls exponentially with the number of gates''~\cite{croke2015pt}, we demonstrate in Sections~\ref{SensingApplication} and~\ref{SearchApplication} that -- even accounting for success probabilities less than one -- this fully unitary embedding still yields substantial practical benefits over standard Hermitian approaches.

We now turn to the technical details of our implementation, starting from the $\mathcal{PT}$-symmetric qubit governed by the Hamiltonian in Eqn.~\eqref{Hamiltonian1}, and then, in Section~\ref{Embed}, showing how it is integrated into a larger Hermitian system, and in Section~\ref{Experimental}, we describe its implementation as a probabilistic quantum gate on the IBM Quantum Experience~\cite{kandala2019error}.

The ket vector both in the Hermitian and $\mathcal{PT}$-symmetric cases has the same form:
\begin{equation}\label{Qubit}
|\psi\rangle = 
\begin{pmatrix}
 \cos\left(\frac{\theta}{2}\right)\\
 e^{i\phi}\sin\left(\frac{\theta}{2}\right)
\end{pmatrix},
\end{equation}
with $\theta$ and $\phi$ being the meridian and parallel of the Bloch sphere of the qubit respectively. The difference lies in the scalar product, which is fixed in the Hermitian case but is \textit{defined} by the $\mathcal{C}$ operator in the $\mathcal{PT}$-symmetric case, as given in Eqn.~\eqref{C_operator}. The scalar product is defined as $\left(\langle\psi|\right)_{\mathcal{CPT}} = \left(\mathcal{CPT}|\psi\rangle\right)^T$ and $\left(\langle\mu|\nu\rangle\right)_{\mathcal{CPT}} = \left(\mathcal{CPT} |\mu\rangle\right)^T \cdot |\nu\rangle$, where the superscript $T$ denotes matrix transposition. The $\mathcal{CPT}$ operation is a combination of the $\mathcal{C}$ and $\mathcal{PT}$ operators, as previously defined. The limit $\alpha \rightarrow 0$ recovers the regular Hermitian case since $\underset{\alpha\rightarrow0}{\lim} \left(\mathcal{C}\right) = \mathcal{P}$.

This property was utilized for $N = 2$ state discrimination~\cite{bender2013observation} to manipulate the angle between state vectors, effectively converting them into orthogonal ones. The no-cloning theorem~\cite{wootters1982single} still applies to both the Hermitian and $\mathcal{PT}$-symmetric cases, as this conversion occurs at the cost of introducing an inconclusive outcome. This means that the $\mathcal{PT}$-symmetric part of the complete wave function of the system in the general case has a norm of less than one. For two non-orthogonal states on the $\phi=-\frac{\pi}{2}$ parallel:
\begin{equation}\label{ReferencePair}
    |\psi_{1,2}\rangle = 
\begin{pmatrix}
\cos\left(\frac{\pi \mp 2\sigma}{4}\right)\\
 -i \sin\left(\frac{\pi \mp 2\sigma}{4}\right)
\end{pmatrix},
\end{equation}
this conversion may be achieved by two possible \textit{Solutions}:

\begin{itemize}
    \item \textit{Solution 1: zeroing the $\mathcal{CPT}$ product, 
     $\left(\mathcal{CPT}\lvert\psi_1\rangle\right)^T\cdot\lvert\psi_2\rangle = 0$, by setting  the Hamiltinian in Eqn.~\eqref{Hamiltonian1} to make $\sin\left(\alpha\right) = \frac{r}{s}\sin\left(\beta\right) =  \cos\left(\sigma\right)$.}
    \item \textit{Solution 2: performing $\mathcal{PT}$-symmetric Hamiltonian evolution to
    $(\langle\psi_1|\psi_2\rangle)_{Hermitian} = 0$ for a time $\tau_{Perp}$ given by:}
    \begin{equation}
\sin^2\left(\omega \tau_{Perp}\right) = \frac{\cos^2\left(\alpha\right) \cos\left(\sigma\right)}{2\sin\left(\alpha\right)\left(1 - \sin\left(\alpha\right)\cos\left(\sigma\right)\right)}, \ \omega = \sqrt{s^2 - r^2\sin^2\left(\beta\right)},
\label{Time}
\end{equation}
\textit{effectively modifying the metrics as:}
    \begin{equation}
\cos^2\left(\alpha\right)e^{+i\mathcal{H}^\dagger t}e^{-i\mathcal{H}t} =
\begin{pmatrix}
\cos^2\left(\omega t - \alpha\right) + \sin^2\left(\omega t\right) & -2i\sin^2\left(\omega t\right)\sin\left(\alpha\right) \\
2i\sin^2\left(\omega t\right)\sin\left(\alpha\right) & \cos^2\left(\omega t + \alpha\right) + \sin^2\left(\omega t\right)
\end{pmatrix}  
\end{equation}
\end{itemize}
In contrast to the Hermitian case, in $\mathcal{PT}$-symmetric dynamics, the states $\ket{0}$ and $\ket{1}$ exhibit an angular separation of $\pi - 2\lvert\alpha\lvert$~\cite{bender2007faster}. As the system approaches the exceptional point in the limit $\alpha\rightarrow\pm\frac{\pi}{2}$, these states merge.

In Section~\ref{ThreeStatesTheory}, we extend the $\mathcal{PT}$-symmetric approach to three states, $N = 3$, through a double $\mathcal{PT}$-symmetric evolution. In Section~\ref{Embed}, we show that for the case of two states, $N = 2$, and minimal value of $\alpha$ allowed by Eqn.~\eqref{Time}, this approach is equivalent to an unambiguous quantum state discrimination~\cite{ivanovic1987differentiate}. In the same Section, we derive the corresponding expressions for $N = 3$. Further in the text, we demonstrate novel features not present in the Hermitian case by leveraging the properties of $\mathcal{PT}$-symmetric exceptional points.

\section{Scheme for $\mathcal{PT}$-symmetric transformation of $N = 3$ states}\label{ThreeStatesTheory}
\subsection{Overview and \textit{Steps} \textit{1} and \textit{2}}

Our approach consists of three steps and can be summarized as follows\footnote{For more details, see: \href{https://github.com/BalytskyiJaroslaw/QuantumSimulations/tree/master}{https://github.com/BalytskyiJaroslaw/QuantumSimulations/tree/master}}:

\begin{itemize}
\item \textit{Step 1}: evolve two of the states,
$\ket{\psi_1}$ and $\ket{\psi_2}$, into the orthogonal ones in terms of the Hermitian scalar product, $(\langle\psi_1|\psi_2\rangle)_{Hermitian} = 0$, by applying the first $\mathcal{PT}$-symmetric evolution.
\item \textit{Step 2}: 
By applying a unitary rotation, convert these effectively orthogonal states into
$\ket{\psi_{1,2}} \rightarrow \frac{1}{\sqrt{2}}\begin{pmatrix}1  \\
\pm i
\end{pmatrix}$. In these positions, they \textit{remain} orthogonal under the $\mathcal{PT}$-symmetric operations with an arbitrary value of the $\alpha$ parameter, allowing us to manipulate this parameter to adjust the relative angle to the third state.

\item \textit{Step 3}: Perform a second $\mathcal{PT}$-symmetric evolution to adjust the angle between $|\psi_{1,2}\rangle$ and $|\psi_3\rangle$, and perform the projective measurement in $\mathcal{PT}$-symmetric subsystem.
\end{itemize}
Without loss of generality, an arbitrary set of three states $\ket{\psi_i} = \begin{pmatrix}\cos\left(\frac{\theta_i}{2}\right)  \\
e^{i\phi_i}\sin\left(\frac{\theta_i}{2}\right)
\end{pmatrix}$, $i \in \left[1, 3\right]$ can be adjusted to the following positions by unitary rotations provided in Eqn.~\eqref{Start} in Methods~\ref{Methods}:
\begin{equation}\label{StartingPosition}
|\psi_{1,2}\rangle \rightarrow \begin{pmatrix}
\cos\left(\frac{\pi \mp 2\sigma}{4}\right)  \\
-i\sin\left(\frac{\pi \mp 2\sigma}{4}\right)
\end{pmatrix},
|\psi_3\rangle \rightarrow \begin{pmatrix}
\cos\left(\frac{\mu}{2}\right)  \\
e^{i\nu}\sin\left(\frac{\mu}{2}\right)
\end{pmatrix},
\end{equation}
and $\sigma$, $\mu$ and $\nu$ parameters in the following equations.

In \textit{Step 1}, we use a $\mathcal{PT}$-symmetric evolution controlled by the Hamiltonian in Eqn.~\eqref{Hamiltonian1}, and perform it for a time $\tau_{Perp}$ in Eqn.~\eqref{Time}. As a result, the first pair of states takes the form:
\begin{equation}
|\psi_1\rangle \rightarrow \begin{pmatrix}
\cos\left(\frac{\delta}{2}\right) \\
-i\sin\left(\frac{\delta}{2}\right)
\end{pmatrix},  \ |\psi_2\rangle \rightarrow \begin{pmatrix}
\sin\left(\frac{\delta}{2}\right)  \\
i\cos\left(\frac{\delta}{2}\right)
\end{pmatrix},
\end{equation}
with the $\delta$ parameter provided by Eqns.~\eqref{Evolved1} and~\eqref{Evolved2}:

\begin{dmath}
\label{Evolved1}
\cos\left(\frac{\delta}{2}\right)= \frac{\cos\left(\omega\tau_{Perp} - \alpha\right)\cos\left(\frac{\pi - 2\sigma}{4}\right)  - \sin\left(\omega\tau_{Perp}\right)\sin\left(\frac{\pi - 2\sigma}{4}\right)}{\sqrt{\mathcal{V}}},
\end{dmath}
\begin{dmath}
\label{Evolved2}
\mathcal{V} = 1 - \cos\left(2\omega\tau_{Perp}\right)\sin^2\left(\alpha\right) +  2\sin\left(\omega\tau_{Perp}\right)\sin\left(\alpha\right) \left(\cos\left(\omega\tau_{Perp}\right)\cos\left(\alpha\right)\sin\left(\sigma\right) - \sin\left(\omega\tau_{Perp}\right)\cos\left(\sigma\right) \right)
\end{dmath}
In \textit{Step 2}, we apply the following gate with the $\chi$ parameter given by Eqn.~\eqref{Step2} in Methods~\ref{Methods}:
\begin{dmath}
\mathcal{W} = 
\frac{1}{\sqrt{2}}
\begin{pmatrix}
1 & i\\
i & 1
\end{pmatrix}\cdot
\begin{pmatrix}
1 & 0\\
0 & ie^{-i\chi}
\end{pmatrix}\cdot
\begin{pmatrix}
\cos\left(\frac{\delta}{2}\right) & i\sin\left(\frac{\delta}{2}\right)\\
i\sin\left(\frac{\delta}{2}\right) & \cos\left(\frac{\delta}{2}\right)
\end{pmatrix},
\end{dmath}
\begin{equation}
\ket{\chi_{\left(1,2,3\right)}} = \mathcal{W}\ket{\psi_{\left(1,2,3\right)}}
\end{equation}
The resulting states take the following form, with $\rho = \xi + \frac{\pi}{2}$ and $\xi$ provided in Eqn.~\eqref{Step2}:
\begin{equation}
\ket{\chi_1} = \frac{1}{\sqrt{2}}\begin{pmatrix}
1  \\
i
\end{pmatrix},  \ \ket{\chi_2} = \frac{1}{\sqrt{2}}\begin{pmatrix}
1  \\
- i
\end{pmatrix},
\ \ket{\chi_3} = \begin{pmatrix}
\cos\left(\frac{\rho}{2}\right)  \\
i\sin\left(\frac{\rho}{2}\right)
\end{pmatrix}
\label{IntermediateAfter12}
\end{equation}
After \textit{Step 2}, the first two states are orthogonal, and we can adjust their relative angles to the third state through the second $\mathcal{PT}$-symmetric evolution, constituting \textit{Step 3}. For completeness, we consider both Hermitian and $\mathcal{CPT}$ measurements corresponding to the aforementioned \textit{Solution 1} and \textit{Solution 2}.

\subsection{\textit{Step 3} by Hermitian measurement}\label{Step3}

\begin{figure}[t]
\begin{minipage}[b]{0.3\linewidth}
\centering
\includegraphics[width=\textwidth, height=5cm]{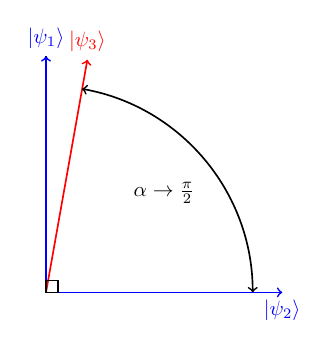}
\caption{Geometry of postselected space for $\alpha\rightarrow\frac{\pi}{2}$.}\label{Fig1}
\end{minipage}
\hspace{0.25cm}
\begin{minipage}[b]{0.3\linewidth}
\centering
\includegraphics[width=\textwidth, height=5cm]{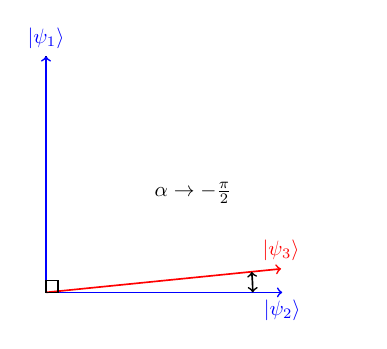}
\caption{Geometry of postselected space for $\alpha\rightarrow-\frac{\pi}{2}$.}
\label{Fig2}
\end{minipage}
\hspace{0.25cm}
\begin{minipage}[b]{0.3\linewidth}
\centering
\includegraphics[width=\textwidth, height=5cm]{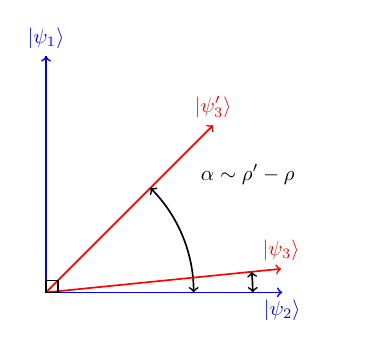}
\caption{Geometry modification by $\alpha$ variation.}
\label{Fig3}
\end{minipage}
\end{figure}

We apply the second $\mathcal{PT}$-symmetric evolution by the Hamiltonian in Eqn.~\eqref{Hamiltonian1} for a time $\tau^{\rom{2}}$, and the relative angles between the evolved states $\kappa_{12}$, $\kappa_{13}$ and $\kappa_{23}$ are given by: 
\begin{dmath}
\begin{cases}
\cos_{\mathcal{PT}}^2\left(\kappa_{12}\right) =
\frac{2 \tan ^2(\alpha ) \sin ^2(2 \omega \tau^{\rom{2}} )}{1 + \sec ^2(\alpha ) - \tan ^2\left(\alpha \right) \cos \left(4\omega \tau^{\rom{2}} \right)} \\
\cos_{\mathcal{PT}}^2\left(\kappa_{13},\kappa_{23}\right) = \\
\frac{\left(\sqrt{2} \sin \left(\frac{\pi \pm 2 \rho}{4} \right) \left[(1 \pm 2 \sin (\alpha )) \sin ^2(\omega \tau^{\rom{2}} )+\cos ^2(\omega \tau^{\rom{2}} + \alpha )\right]+\sin (2
   \alpha ) \cos \left(\frac{\rho }{2}\right) \sin (2 \omega \tau^{\rom{2}} )\right)^2}{2 \left((1\pm\sin (\alpha ))^2 \sin ^2(\omega \tau^{\rom{2}} )+\cos ^2(\alpha ) \cos ^2(\omega
   \tau^{\rom{2}} )\right) \left(\sin ^2(\omega \tau^{\rom{2}} ) (1 + 2 \sin (\alpha ) \sin (\rho ))-\sin (2 \alpha ) \sin ^2\left(\frac{\rho }{2}\right) \sin (2 \omega \tau^{\rom{2}} )+\cos
   ^2(\omega \tau^{\rom{2}} - \alpha)\right)}
\end{cases}
\end{dmath}
By the subscript $\mathcal{PT}$ in $\cos^2_{\mathcal{PT}}$, we mean the effective cosine squared in $\mathcal{PT}$-symmetric subspace after postselection. We derive exact expressions for the maximum probability of successful postselection in the next Section~\ref{Embed}. After the time $\tau^{\rom{2}} = \frac{\pi}{2\omega}$, these expressions take the form:
\begin{equation}\label{Cosines}
\begin{cases}
\cos_{\mathcal{PT}}^2\left(\kappa_{12}\right) = 0 \\
 \cos_{\mathcal{PT}}^2\left(\kappa_{13}, \kappa_{23}\right) = 
\frac{(1 \pm \sin \left(\alpha \right))^2 (1 \pm \sin \left(\rho \right))}{3 + 4 \sin \left(\alpha \right) \sin \left(\rho \right)-\cos \left(2 \alpha \right)}
\end{cases},
\end{equation}
and the normalized state of the qubit is:

\begin{align}
    e^{-i\mathcal{H}\tau^{\rm{II}}}\ket{\chi_3}\Big|_{\tau^{\rm{II}} = \frac{\pi}{2\omega}} 
    &= \frac{1}{\sqrt{1 + 2\sin\alpha\,\sin\rho + \sin^2\alpha}}
    \begin{pmatrix}
        \sin\left(\frac{\rho}{2}\right) + \sin\alpha\,\cos\left(\frac{\rho}{2}\right) \\
        -i\left(\cos\left(\frac{\rho}{2}\right) + \sin\alpha\,\sin\left(\frac{\rho}{2}\right)\right)
    \end{pmatrix}
\end{align}

In the limit $\alpha \rightarrow \frac{\pi}{2}$, we obtain: 

\begin{equation}\label{Fig1Eqn1}
\cos_{\mathcal{PT}}^2\left(\kappa_{13}\right) = 1 - \frac{\left(1 - \sin\left(\rho\right)\right)\left(\frac{\pi}{2} - \alpha\right)^4}{16\left(1 + \sin\left(\rho\right)\right)} + O\left(\left(\frac{\pi }{2} - \alpha\right)^5\right),
\end{equation}
\begin{equation}\label{Fig1Eqn2}
\cos_{\mathcal{PT}}^2\left(\kappa_{23}\right) =  \frac{\left(1 - \sin\left(\rho\right)\right)\left(\frac{\pi}{2} - \alpha\right)^4}{16\left(1 + \sin\left(\rho\right)\right)} + O\left(\left(\frac{\pi }{2} - \alpha\right)^5\right),
\end{equation}
and for $\alpha \rightarrow -\frac{\pi}{2}$, we have: 

\begin{equation}\label{Fig2Eqn1}
\cos_{\mathcal{PT}}^2\left(\kappa_{13}\right) = \frac{(1 + \sin (\rho ))\left(\frac{\pi }{2} + \alpha \right)^4 }{16\left(1- \sin (\rho )\right)}+O\left(\left(\frac{\pi }{2} + \alpha\right)^5\right),
\end{equation}

\begin{equation}\label{Fig2Eqn2}
\cos_{\mathcal{PT}}^2\left(\kappa_{23}\right) = 1 -  \frac{(1 + \sin (\rho ))\left(\frac{\pi }{2} + \alpha \right)^4 }{16\left(1- \sin (\rho )\right)}+O\left(\left(\frac{\pi }{2} + \alpha\right)^5\right)
\end{equation}
The corresponding geometry of the states in these limits is shown in Fig.~\ref{Fig1} and Fig.~\ref{Fig2}. These results \textit{apparently} seem contradictory to the well-known impossibility of unambiguous discrimination of linearly dependent states~\cite{chefles1998unambiguous}. However, such $\mathcal{PT}$-symmetric transformation inevitably involves postselection, and in the next Section~\ref{Embed}, we show that changing the angles in the $\mathcal{PT}$-symmetric subspace happens at the cost of reduction of probability of successful postselection. When considering the probability of postselection, these results align with those in~\cite{chefles1998unambiguous}, as we show further in the text.

As we showed earlier, an arbitrary set of three states can be reduced to the states in Eqn.~\eqref{IntermediateAfter12} through $\mathcal{PT}$-symmetric transformations. Thus, an arbitrary set of three states is uniquely characterized by its $\rho$ value, up to the initial unitary transformation described in Methods~\ref{Methods}. Therefore, the parameter $\alpha$ can be used to convert a set of three states characterized by the parameter $\rho$ into another set of three states corresponding to the parameter $\rho^\prime$. This can be done by setting the value of $\alpha$ to be:

\begin{equation}\label{Transformer}
\sin\left(\alpha\right) = \Big\{
\frac{\cos \left(\frac{\rho + \rho^\prime }{2}\right)}{\sin
   \left(\frac{\rho^\prime - \rho}{2}\right)} ,
   \frac{\sin\left(\frac{\rho^\prime - \rho}{2}\right)}{\cos \left(\frac{\rho + \rho^\prime}{2}\right)}
\Big\},
\end{equation}
depending on the values of $\rho$ and $\rho^\prime$ to ensure that $\lvert\sin\left(\alpha\right)\lvert\le 1$, as illustrated in Fig.~\ref{Fig3}. By running these \textit{Steps} backward, one can transform the second set of states back to the first set. As we discuss in Section~\ref{Embed}, when the postselection probability is taken into account, this does not lead to a reduction in error rate compared to conventional Hermitian approaches. However, this property may be useful for the discrimination of states with highly asymmetric geometries, as we discuss in Section~\ref{TrineQKD}.

As an example, one can achieve an \textit{effective} mirror-symmetric geometry of postselected states corresponding to $\rho^\prime = 0$, when
$\cos_{\mathcal{PT}}^2\left(\kappa_{13}\right) = \cos_{\mathcal{PT}}^2\left(\kappa_{23}\right) = \frac{1}{2}$ in Eqn.~\eqref{Cosines}, by choosing:

\begin{equation}\label{MirrorCondition}
\sin\left(\alpha\right) = \Big\{-\cot\left(\frac{\rho}{2}\right),  -\tan\left(\frac{\rho}{2}\right)\Big\}
\end{equation}
After applying the $S$ gate:
\begin{equation}
S = \begin{pmatrix}
1 & 0 \\
0 & i
\end{pmatrix},
\end{equation}
this set of states is transformed to $\ket{+}$, $\ket{-}$, and $\ket{0}$, which are stabilizer states~\cite{gottesman1997stabilizer}. As we show in Section~\ref{Embed}, even though the \textit{effective} geometry of postselected states is mirror-symmetric, the postselection changes the prior probabilities. Therefore, even though the effective angles $\kappa_{13} = \kappa_{23} = \frac{\pi}{4}$ are the same, the prior probabilities of $\ket{\psi_1}$ and $\ket{\psi_2}$ are different in the general case.

\subsection{\textit{Step 3} by \texorpdfstring{$\mathcal{CPT}$}{CPT} measurement}

The same result can be achieved using the $\mathcal{CPT}$ measurement, since for an arbitrary $\alpha$, the states $\ket{\psi_1}$ and $\ket{\psi_2}$ are mutually orthogonal:
\begin{equation}
\left(\langle\psi_1|\psi_2\rangle \right)_{\mathcal{CPT}} = 0; \
\left(\bra{\psi_{1,2}}\right)_{\mathcal{CPT}} = \frac{\left(1 \pm \sin\left(\alpha\right)\right)}{\sqrt{2}\cos\left(\alpha\right)}
\left(
1, \\
\mp i
\right)
\end{equation}
This allows the value of $\alpha$ to be used to adjust the relative angles to the third state, $\kappa_{13}$ and $\kappa_{23}$:
\begin{equation}
\begin{cases}
\cos_{\mathcal{PT}}^2\left(\kappa_{12}\right) = 0 \\
\cos_{\mathcal{PT}}^2\left(\kappa_{13},\kappa_{23}\right) = \frac{\left(1 \pm \sin\left(\alpha\right)\right)\left(1 \pm \sin\left(\rho\right)\right)}{2\left(1 + \sin\left(\alpha\right)\sin\left(\rho\right)\right)}
\end{cases},
\end{equation}
and for $\alpha \rightarrow \frac{\pi}{2}$ represented in Fig.~\ref{Fig1}:
\begin{equation}
\cos_{\mathcal{PT}}^2\left(\kappa_{13}\right)  = 1 - \frac{\left(\frac{\pi}{2} - \alpha\right)^2 \left(1-\sin (\rho )\right)}{4 (1+\sin (\rho ))} + O\left(\left(\frac{\pi }{2} -\alpha \right)^3\right),
\end{equation}

\begin{equation}
\cos_{\mathcal{PT}}^2\left(\kappa_{23}\right)  = \frac{\left(\alpha -\frac{\pi}{2}\right)^2 \left(1-\sin (\rho )\right)}{4 (1+\sin (\rho ))} + O\left(\left(\alpha -\frac{\pi }{2}\right)^3\right)
\end{equation}
In the limit $\alpha \rightarrow -\frac{\pi}{2}$ corresponding to Fig.~\ref{Fig2}:
\begin{equation}
\cos_{\mathcal{PT}}^2\left(\kappa_{13}\right)  = \frac{\left(\frac{\pi }{2} + \alpha\right)^2 \left(1+\sin (\rho )\right)}{4 (1-\sin (\rho ))} + O\left(\left(\frac{\pi }{2} + \alpha\right)^3\right),
\end{equation}

\begin{equation}
\cos_{\mathcal{PT}}^2\left(\kappa_{23}\right)  = 1 - \frac{\left(\frac{\pi }{2} + \alpha\right)^2 \left(1+\sin (\rho )\right)}{4 (1-\sin (\rho ))} + O\left(\left(\frac{\pi }{2} + \alpha\right)^3\right)
\end{equation}
Analogously to the Hermitian case, the $\mathcal{CPT}$ projection operators can be introduced, which are the $\mathcal{CPT}$ observables:

\begin{equation}
\hat{P}_{1,2} = \left(\frac{|\psi_{1,2}\rangle\langle\psi_{1,2}|}{\langle\psi_{1,2}|\psi_{1,2}\rangle}\right)_{\mathcal{CPT}}=  \frac{1}{{2}}
\begin{pmatrix}
1 & \mp i\\
\pm i & 1
\end{pmatrix}, \left[\mathcal{CPT}, \hat{P}_{1, 2}\right] = 0
\end{equation}
Similarly to the Hermitian case in Eqn.~\eqref{Transformer}, it is possible to transform the states $\rho\rightarrow\rho^\prime$ by choosing:

\begin{equation}
    \sin\left(\alpha\right) = \frac{\sin\left(\rho^\prime\right) - \sin\left(\rho\right) }{1 - \sin\left(\rho^\prime\right)\sin\left(\rho\right)},
\end{equation}
and when $\alpha=-\rho$, three states are reduced to effectively mirror-symmetric corresponding to $\rho^\prime = 0$, as illustrated in Fig.~\ref{Fig3}.

Unlike prior studies that relied on numerical computations~\cite{wu2019observation, dogra2021quantum}, in the next Section~\ref{Embed}, we derive \textit{precise} expressions for the maximum probability of achieving a definitive outcome following $\mathcal{PT}$-symmetric evolution. This approach allows for a direct and fair comparison of $\mathcal{PT}$-symmetric systems with their Hermitian counterparts in the following Sections.

\section{Embedding by the dilation  method for $N = 2$ and $N = 3$}\label{Embed}

We implement the $\mathcal{PT}$-symmetric Hamiltonian evolution by extending the original qubit with ancilla and employing Neumark's theorem~\cite{neumark}, similarly to~\cite{wu2019observation, dogra2021quantum}. The combined ancilla-qubit wave function $\ket{\Psi_{combined}\left(t\right)}$ with the $\mathcal{PT}$-symmetric subspace $\ket{\psi_{\mathcal{PT}}\left(t\right)}$ is:

\begin{equation}
\label{CombinedStateWorkAncilla}
    \ket{\Psi_{combined}\left(t\right)} = \ket{\psi_{\mathcal{PT}}\left(t\right)}\ket{0}_{ancilla} + \zeta\left(t\right)\ket{\psi_{\mathcal{PT}}\left(t\right)}\ket{1}_{ancilla},
\end{equation}
where the operator $\zeta\left(t\right) = \zeta^\dagger\left(t\right) = \left(\mathcal{N}\left(t\right) - \hat{1}\right)^{\frac{1}{2}}$ must maintain its eigenvalues real throughout the entire duration of the $\mathcal{PT}$-symmetric evolution. The initial value $\mathcal{N}\left(0\right)$ must be correspondingly chosen in order to ensure it, with:
 
\begin{equation}
\mathcal{N}\left(t\right) = T \exp\left[-i\int_0^t\mathrm{d}\tau^\prime \ \mathcal{H}^\dagger_q\left(\tau^\prime\right)\right]\mathcal{N}\left(0\right)\tilde{T}\exp\left[i\int^t_0 \mathrm{d}\tau^\prime \ \mathcal{H}\left(\tau^\prime\right)\right],
\end{equation}
where $T$ and $\tilde{T}$ are the time and and anti-time-ordering operators, respectively. 

Further in the text, we perform analytical computations to find the minimal value of $\mathcal{N}\left(0\right)$ that maximizes the probability of the conclusive outcome. Thus, unlike numerical computations in~\cite{wu2019observation, dogra2021quantum},
for both the first and second stages of $\mathcal{PT}$-symmetric evolution, we find the population of postselected space \textit{exactly} by simplifying the following equation:

\begin{equation}\label{Decisiveness}
    \mathcal{D} = \frac{\langle\psi_{\mathcal{PT}}\left(t\right)\ket{\psi_{\mathcal{PT}}\left(t\right)}}{\langle\psi_{\mathcal{PT}}\left(t\right)\ket{\psi_{\mathcal{PT}}\left(t\right)} + \bra{\psi_{\mathcal{PT}}\left(t\right)}\zeta^2\left(t\right)\ket{\psi_{\mathcal{PT}}\left(t\right)}}, 
\end{equation}

\subsection{First stage, and $N = 2$ case}

As demonstrated experimentally in~\cite{chen2022quantum, wang2024demonstration}, at the critical value, $\mathcal{PT}$-symmetric quantum state discrimination has the same success probability as the optimal unambiguous state discrimination in Hermitian systems~\cite{ivanovic1987differentiate}. Our present work provides the analytical derivation that was previously lacking in the literature.

First, the smallest value of $\alpha$ in Eqn.~\eqref{Time} allowing to perform $\mathcal{PT}$-symmetric evolution corresponding to $\sin^2\left(\omega \tau_{Perp}\right)  = 1$ is given by:

\begin{equation}\label{Critical}
\sin\left(\alpha\right) = (1-\sin\left(\sigma\right)) \sec\left(\sigma\right)
\end{equation}
For added convenience, alongside the pair of reference vectors in Eqn.~\eqref{ReferencePair}, we introduce the vector situated between them, aligning along the same parallel of the Bloch sphere:

\begin{equation}
    |\psi_{m}\rangle = 
\begin{pmatrix}
\cos\left(\frac{\pi + 2 m}{4}\right)\\
 -i \sin\left(\frac{\pi + 2 m}{4}\right)
\end{pmatrix}
\label{TwoProbeState}
\end{equation}
The resulting $\cos^2_{\mathcal{PT}}\left(\ket{\psi_m},\ket{\psi_1}\right)$ in the postselected subspace turns out to be the same as computed by the $\mathcal{CPT}$ scalar product ~\cite{balytskyi2021detecting}:
\begin{equation}
\cos^2_{\mathcal{PT}}\left(\ket{\psi_m},\ket{\psi_1}\right) = \frac{1 - \cos \left(m - \sigma \right)}{2\left(1 - \cos \left(m \right) \cos \left(\sigma \right)\right)}
\label{CosTwoProbeState}
\end{equation}

By explicitly computing the eigenvalues of $\zeta\left(t\right)$, we find that the requirement that they remain real throughout the evolution simplifies to the condition:
\begin{equation}\label{ConditionReal1}
\mathcal{N}\left(0\right) \cot \left(\frac{\sigma }{2}\right)-1 \ge 0 \ \ \& \ \ \mathcal{N}\left(0\right) \tan \left(\frac{\sigma }{2}\right)-1 \ge 0,
\end{equation}
and thus:
\begin{equation}
  \mathcal{N}\left(0\right)  = max\Big\{\tan\left(\frac{\sigma }{2}\right), \cot \left(\frac{\sigma }{2}\right)\Big\}
\end{equation}
For $0 < \sigma < \frac{\pi}{2}$ and $\cos\left(\sigma\right) > 0$, one needs to choose $\mathcal{N}\left(0\right)  = \cot\left(\frac{\sigma }{2}\right)$, and at the end of the $\mathcal{PT}$-symmetric evolution, the $\zeta$ operator can be found explicitly:
\begin{equation}
\zeta^{\rm{I}}_+\left(\tau^{\rm{I}} = \frac{\pi}{2\omega}\right) = \frac{1}{2} \sqrt{\cos (\sigma )} \csc \left(\frac{\sigma }{2}\right)\left(
\begin{array}{cc}
 1 & -i \\
 i & 1 \\
\end{array}
\right),
\end{equation}
as well as the population of postselected space representing the probability of a conclusive outcome:
\begin{equation}
\mathcal{D}^{\rm{I}}_+\left(m,\sigma\right) = \frac{1}{2}\left(1 -\cos\left(m\right)\cos\left(\sigma\right)\right)\sec^2\left(
\frac{\sigma}{2}\right)
\end{equation}
On the edges corresponding to $\ket{\psi_1}$ and $\ket{\psi_2}$, one finds:
\begin{equation}
\label{DPCase1}
\mathcal{D}^{\rm{I}}_+\left(\sigma,\sigma\right) = \mathcal{D}^{\rm{I}}_+\left(- \sigma,\sigma\right) = 1 - \cos\left(\sigma\right) =  1 - \lvert\cos\left(\sigma\right)\lvert
\end{equation}
For $ \frac{\pi}{2} < \sigma < \pi$ and $\cos\left(\sigma\right) < 0$, one needs to choose $\mathcal{N}\left(0\right)  = \tan\left(\frac{\sigma }{2}\right)$ leading to the following $\zeta$ operator at the end of evolution:
\begin{equation}
\zeta^{\rm{I}}_-\left(\tau^{\rm{I}} = \frac{\pi}{2\omega}\right) = \frac{1}{2}\sqrt{-\cos (\sigma )} \sec \left(\frac{\sigma }{2}\right)\left(
\begin{array}{cc}
 1 & i \\
 -i & 1 \\
\end{array}
\right),
\end{equation}
and the corresponding population of postselected space is:
\begin{equation}
\mathcal{D}^{\rm{I}}_-\left(m,\sigma\right) = \frac{1}{2}\left(1 -\cos\left(m\right)\cos\left(\sigma\right)\right)\csc^2\left(
\frac{\sigma}{2}\right)
\label{ThreeQKDDec},
\end{equation}
\begin{equation}
\label{DPCase2}
\mathcal{D}^{\rm{I}}_-\left(\sigma,\sigma\right) = \mathcal{D}^{\rm{I}}_-\left(- \sigma,\sigma\right) = 1 + \cos\left(\sigma\right) =  1 - \lvert\cos\left(\sigma\right)\lvert
\end{equation}
Combining Eqns.~\eqref{DPCase1} and~\eqref{DPCase2}, one observes that, when probability of successful outcome is considered, the $\mathcal{PT}$-symmetric discrimination of $N = 2$ quantum states developed in~\cite{bender2013pt} converts two reference vectors in Eqn.~\eqref{ReferencePair} to orthogonal ones with the probability of the conclusive outcome being $1 - \lvert\cos\left(\sigma\right)\lvert$, the same as in a conventional unambiguous quantum state discrimination~\cite{ivanovic1987differentiate}. We extend this result for $N = 3$ states in the next Subsection.

\begin{figure}[h!]
\includegraphics[width=\textwidth]{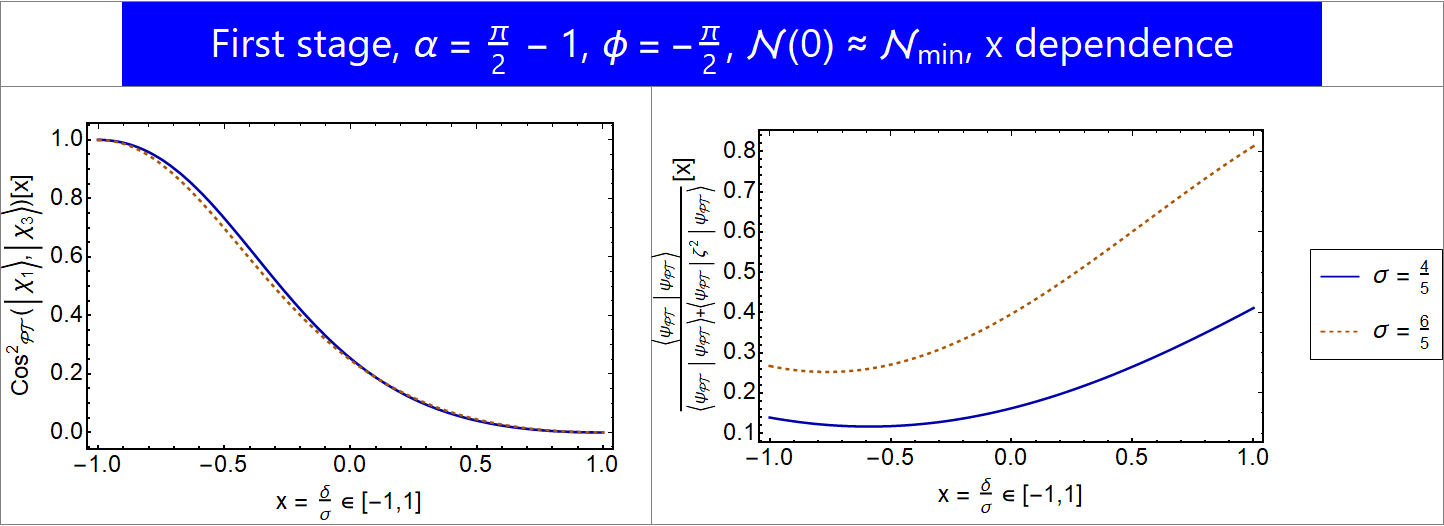}
\caption{$\cos^2_{\mathcal{PT}}$ in $\mathcal{PT}$-symmetric subsystem and population of postselected space corresponding to \textit{Stage 1} of our algorithm, using expressions from Section~\ref{Embed}, corresponding to the input state $\ket{\psi_3}$ as defined in Eqn.~\eqref{ProbeStates}, and subsequently implemented on IBM Quantum Experience in Section~\ref{Experimental}.}
\label{FirstStage1}
\end{figure}

\begin{figure}[h!]
\includegraphics[width=\textwidth]{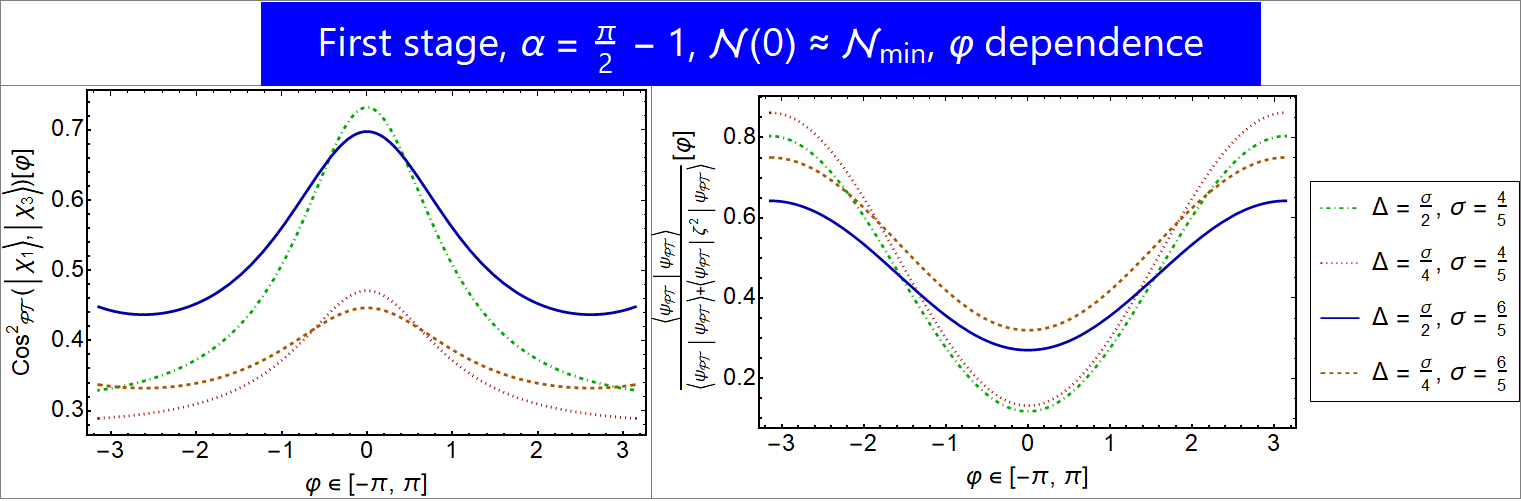}
\caption{$\cos^2_{\mathcal{PT}}$ in $\mathcal{PT}$-symmetric subsystem and population of postselected space corresponding to \textit{Stage 1} of our algorithm, using expressions from Section~\ref{Embed}, corresponding to the input state $\ket{\psi^\prime_3}$ as defined in Eqn.~\eqref{ProbeStates}, and subsequently implemented on IBM Quantum Experience in Section~\ref{Experimental}.}
\label{FirstStage2}
\end{figure}
Figs.~\ref{FirstStage1} and~\ref{FirstStage2} show several representative values of parameters used as input for \textit{Stage 1}. It can be observed that higher values of $\sigma$, representing quantum states with better distinguishability, correspond to higher probabilities of a decisive outcome. Using these parameter values, we implement \textit{Stage 1} on IBM Quantum Experience in the next Section~\ref{Experimental}.

\subsection{Second stage, and $N = 3$ case}
\begin{figure}[h!]
\includegraphics[width=\textwidth]{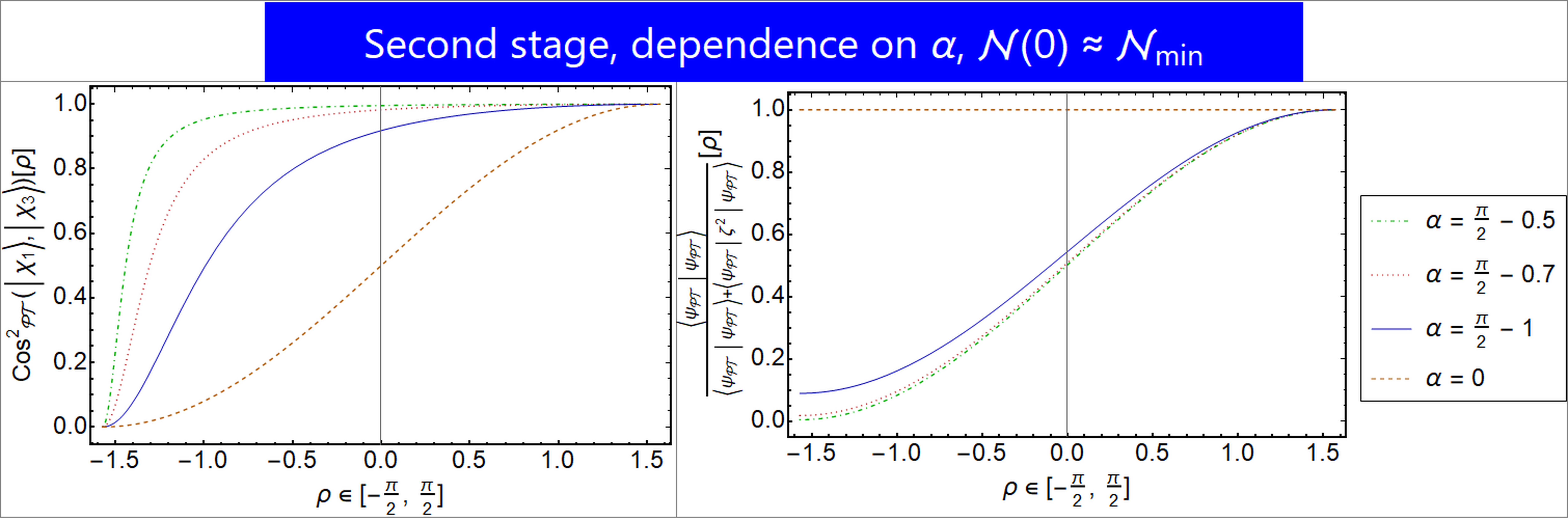}
\caption{$\cos^2_{\mathcal{PT}}$ in $\mathcal{PT}$-symmetric subsystem and population of postselected space for  \textit{Stage 2} of our algorithm with $\alpha > 0$, using expressions from Section~\ref{Embed}, corresponding to the input state $\ket{\chi_3}$ as defined in Eqn.~\eqref{IntermediateAfter12}, and subsequently implemented on IBM Quantum Experience in Section~\ref{Experimental}.}
\label{SecondStageTh}
\end{figure}

Similarly to \textit{Stage 1}, by explicitly computing the $\zeta$ operator for the case of \textit{Stage 2}, we find that the condition on the reality of its eigenvalues reduces to:

\begin{equation}
\small
\mathcal{N}\left(0\right) \ge \frac{1 + \cos (2 \alpha )}{2 -\cos (2 \omega \tau^{\rm{II}} ) + \cos (2 \alpha ) \cos (2 \omega \tau^{\rm{II}}) \pm 2 \sin (\alpha ) \sin (\omega \tau^{\rm{II}}) \sqrt{3+\cos (2 \alpha ) -2 \sin ^2(\alpha ) \cos (2 \omega \tau^{\rm{II}})}}\label{SecondStageInit}
\end{equation}
For $0 \le \alpha < \frac{\pi}{2}$, one needs to choose the ``-'' sign which leads to the following $\zeta$ operator at the end of evolution:

\begin{equation}
\zeta^{\rm{II}}_+\left(\tau^{\rm{II}} = \frac{\pi}{2\omega}\right)  =\frac{\sqrt{\sin (\alpha )}}{1-\sin (\alpha )}\left(
\begin{array}{cc}
 1 & -i \\
 i & 1 \\
\end{array}
\right)
\end{equation}
This leads to the following probability of a decisive outcome:
\begin{equation}
\label{Dec1}
\mathcal{D}^{\rm{II}}_+\left(\alpha,\rho\right) = \frac{3 + 4 \sin (\alpha ) \sin (\rho )-\cos (2 \alpha )}{3 + 4 \sin (\alpha )-\cos (2 \alpha )}
\end{equation}
Observe that
$\mathcal{D}^{\rm{II}}_+\left(\alpha,\rho = \frac{\pi}{2}\right) = 1$, and in the limit $\alpha\rightarrow\frac{\pi}{2}$:

\begin{equation}
\mathcal{D}^{\rm{II}}_+\left(\alpha,\rho\right) = \frac{1}{2} (1 + \sin (\rho ))+\frac{1}{32} \left(\frac{\pi }{2} -\alpha\right)^4 (1-\sin (\rho
   ))+O\left(\left(\frac{\pi }{2} -\alpha\right)^5\right)
\end{equation}
Thus, in this limit, the position $\rho = -\frac{\pi}{2}$ becomes close to almost always inconclusive.

Similarly, for $- \frac{\pi}{2} < \alpha \le 0$, one chooses the ``+'' in Eqn.~\eqref{SecondStageInit} and obtains:

\begin{equation}
\zeta^{\rm{II}}_-\left(\tau^{\rm{II}} = \frac{\pi}{2\omega}\right)  = \frac{\sqrt{-\sin (\alpha )}}{1 + \sin (\alpha )}\left(
\begin{array}{cc}
 1 & i \\
 -i & 1 \\
\end{array}
\right),
\end{equation}
\begin{equation}
\mathcal{D}^{\rm{II}}_-\left(\alpha, \rho\right) = \frac{3 + 4 \sin (\alpha ) \sin (\rho ) - \cos (2 \alpha )}{3 - 4 \sin (\alpha ) - \cos (2 \alpha )}\label{Dec2},
\end{equation}
and for this case, $\mathcal{D}^{\rm{II}}_-\left(\alpha,\rho = -\frac{\pi}{2}\right) = 1$, and in the limit $\alpha\rightarrow-\frac{\pi}{2}$:

\begin{equation}
\mathcal{D}^{\rm{II}}_-\left(\alpha, \rho\right) = \frac{1}{2} (1-\sin (\rho ))+\frac{1}{32} \left(\alpha +\frac{\pi }{2}\right)^4 (1 + \sin (\rho ))+O\left(\left(\alpha +\frac{\pi }{2}\right)^5\right)
\end{equation}
Importantly, combining two cases $\mathcal{D}^{\rm{II}}_{\pm}$, we observe that:

\begin{equation}
\label{Leverage1}
\cos_{\mathcal{PT}}^2\left(\kappa_{23}\right)\mathcal{D}^{\rm{II}}_+\left(\alpha,\rho\right) = \frac{(1-\sin (\alpha ))^2 (1-\sin (\rho ))}{3 + 4 \sin (\alpha )-\cos
   (2 \alpha )},
\end{equation}

\begin{equation}
\label{Leverage2}
\cos_{\mathcal{PT}}^2\left(\kappa_{13}\right)\mathcal{D}^{\rm{II}}_-\left(\alpha,\rho\right) = \frac{(1 + \sin (\alpha ))^2 (1 + \sin (\rho ))}{3-4 \sin (\alpha
   )-\cos (2 \alpha )}
\end{equation}
And similarly:

\begin{equation}
\label{Leverage3}
\cos_{\mathcal{PT}}^2\left(\kappa_{13}\right)\mathcal{D}^{\rm{II}}_+\left(\alpha,\rho\right) = \frac{1 + \sin\left(\rho\right)}{2} = \cos^2\left(\frac{\frac{\pi}{2} - \rho }{2}\right),
\end{equation}

\begin{equation}
\label{Leverage4}
\cos_{\mathcal{PT}}^2\left(\kappa_{23}\right)\mathcal{D}^{\rm{II}}_-\left(\alpha,\rho\right) = \frac{1 - \sin\left(\rho\right)}{2} = \cos^2\left(\frac{\frac{\pi}{2} + \rho }{2}\right)
\end{equation}
From Eqns.~\eqref{Leverage1} and~\eqref{Leverage2}, it can be observed that when the probability of a decisive outcome is taken into account, the $\mathcal{PT}$-symmetric transformation during \textit{Stage 2} does \underline{not} improve state distinguishability. The state at the $\mathcal{PT}$-symmetric exceptional point ($\ket{\chi_1}$ or $\ket{\chi_2}$) has a low probability of a conclusive outcome. The state $\ket{\chi_3}$, upon successful postselection, has a small projection on the reference vector corresponding to the exceptional point. However, its decisiveness (the probability of successful postselection) is much higher than that of the state corresponding to the exceptional point ($\ket{\chi_1}$ or $\ket{\chi_2}$), as observed in Eqns.~\eqref{Dec1} and~\eqref{Dec2}. As a result, the average error rate remains the same as in the Hermitian case since all values of $\rho$ in Eqns.~\eqref{Leverage1} and~\eqref{Leverage2} are rescaled by the same factor. Similarly, from Eqns.~\eqref{Leverage3} and~\eqref{Leverage4}, one observes that an increase in $\cos_{\mathcal{PT}}^2$ is accompanied by a reduction of $\mathcal{D}^{\rm{II}}_{\pm}$, resulting in the same average outcome as in the Hermitian case.

Nevertheless, as we show in Section~\ref{SensingApplication}, the ability to consolidate all relevant information about the parameter of interest within a small subset of events provides significant technical advantages over conventional Hermitian systems in terms of the information-cost rate. Additionally, as demonstrated in Section~\ref{SearchApplication}, Eqn.~\eqref{Leverage3} indicates that the application of $\mathcal{PT}$-symmetric operations does \underline{not} reduce the number of oracle calls needed to find a target in the unstructured quantum database. However, when combined with the punctuated version of Grover's search algorithm, it significantly reduces the \textit{qubit readout cost}, presenting substantial technical advantages for NISQ computers.

Additionally, observe that for any value of $\alpha$, a failed post-selection event -- corresponding to the ancilla being measured as $\ket{1}_{ancilla}$ in Eqn.~\eqref{CombinedStateWorkAncilla} -- deterministically collapses the qubit state into a form that is independent of $\rho$, up to normalization:

\begin{align}
\label{ZetaPlusConst}
&\zeta^{\rm{II}}_+\left(\tau^{\rm{II}}\right)\, e^{-i\mathcal{H}\tau^{\rm{II}}} \ket{\chi_3} \Big|_{\tau^{\rm{II}} = \frac{\pi}{2\omega}} 
= \frac{\sqrt{\sin\alpha}}{1 - \sin\alpha}
\begin{pmatrix}
 1 & -i \\
 i & 1
\end{pmatrix}
\cdot \sec\alpha
\begin{pmatrix}
 \sin\alpha & -i \\
 -i & -\sin\alpha
\end{pmatrix}
\begin{pmatrix}
\cos\left(\frac{\rho}{2}\right) \\
i \sin\left(\frac{\rho}{2}\right)
\end{pmatrix} \nonumber \\
&\qquad = -\frac{\sqrt{2 \sin\alpha}}{\cos\alpha} \cos\left( \frac{\rho}{2} + \frac{\pi}{4} \right)
\begin{pmatrix}
1 \\
i
\end{pmatrix}, \quad \text{for } 0 \le \alpha < \frac{\pi}{2},
\end{align}

\begin{align}
\label{ZetaMinusConst}
&\zeta^{\rm{II}}_-\left(\tau^{\rm{II}}\right)\, e^{-i\mathcal{H}\tau^{\rm{II}}} \ket{\chi_3} \Big|_{\tau^{\rm{II}} = \frac{\pi}{2\omega}} 
= \frac{\sqrt{-\sin\alpha}}{1 + \sin\alpha}
\begin{pmatrix}
 1 & i \\
 -i & 1
\end{pmatrix}
\cdot \sec\alpha
\begin{pmatrix}
 \sin\alpha & -i \\
 -i & -\sin\alpha
\end{pmatrix}
\begin{pmatrix}
\cos\left(\frac{\rho}{2}\right) \\
i \sin\left(\frac{\rho}{2}\right)
\end{pmatrix} \nonumber \\
&\qquad = \frac{\sqrt{-2 \sin\alpha}}{\cos\alpha} \sin\left( \frac{\rho}{2} + \frac{\pi}{4} \right)
\begin{pmatrix}
1 \\
-i
\end{pmatrix}, \quad \text{for } -\frac{\pi}{2} < \alpha \le 0.
\end{align}
In particular, as discussed in Section~\ref{SensingApplication}, whenever post-selection fails, the Quantum Fisher Information is zero, since the resulting density matrix is constant in this case.

Furthermore, in the limit $\alpha \rightarrow \pm \frac{\pi}{2}$, the $\mathcal{PT}$-symmetric transformation described in Subsection~\ref{Step3} maps the qubit -- initially encoding an \emph{unknown} parameter $\rho$ -- into a known pure state, conditioned on the binary outcome of an ancilla measurement. If post-selection succeeds in this limit, the qubit deterministically approaches the fixed state, as $\cos^2_{\mathcal{PT}} \rightarrow 1$ in Eqns.~\eqref{Fig1Eqn1} and~\eqref{Fig2Eqn2}. Conversely, if post-selection fails, the qubit state becomes fully determined, as shown in Eqns.~\eqref{ZetaPlusConst} and~\eqref{ZetaMinusConst}. We leverage this property in Section~\ref{SearchApplication}, applying it to the punctuated version of Grover's search algorithm, and performing the final measurement of all working qubits only when post-selection succeeds -- thus reducing the qubit measurement cost in depth-limited quantum searches.

For $\alpha > 0$, the output of \textit{Stage 2} is illustrated  Fig.~\ref{SecondStageTh}, and in the next Section~\ref{Experimental},
we confirm these analytical results by numerical computations and simulations on IBM Quantum Experience.

\section{Implementation on IBM Quantum Experimence}\label{Experimental}

IBM Quantum Experience~\cite{kandala2019error} is a quantum processor operating on superconducting qubits that has become a leading candidate for scalable quantum computing platform, see a review~\cite{huang2020superconducting}. These devices already enabled proof-of-concept results such as quantum error correction~\cite{devitt2016performing}, fault-tolerant gates~\cite{harper2019fault}, experimental evidence of the violation of Mermin and Leggett-Garg inequalities~\cite{alsina2016experimental,huffman2017violation}, non-local parity
measurements~\cite{hegade2019investigation,paraoanu2018non}, simulations of paradigmatic models in open
quantum systems~\cite{garcia2020ibm}, creation of highly entangled graph states~\cite{wang201816}, determining the ground-state energies of the molecules~\cite{kandala2017hardware} as well as implementation of quantum witnesses~\cite{ku2020experimental}. Moreover, $\mathcal{PT}$-symmetric quantum mechanics can enhance entanglement by local operations, a possibility prohibited in the Hermitian case, as demonstrated experimentally by IBM Quantum Experience~\cite{dogra2021quantum} based on theoretical findings from~\cite{chen2014increase}.

To verify our analytical results in Section~\ref{Embed}, we implemented both \textit{Stage 1} and \textit{Stage 2} on IBM Quantum Experience, with details on numerical inputs provided in Methods~\ref{Methods}. We performed experiments on all processors provided by IBM Quantum Experience, namely ibm\_perth, ibmq\_jakarta, ibm\_lagos, ibm\_nairobi, ibm\_oslo, ibmq\_manila, ibmq\_quito, ibmq\_belem, ibmq\_lima, simulator\_mps, simulator\_extended\_stabilizer, \\
ibmq\_qasm\_simulator, simulator\_statevector. In each experiment, the total number of shots was kept $N_{shots}=8192$, and $N\left(\ket{ij}\right)$, $i\in\left[0,1\right]$ is a number of outcomes corresponding to $\ket{ij}$, such that $\sum_{\substack{i,j = 1,2}}$ $N\left(\ket{ij}\right) = N_{shots}$.

For both \textit{Stages}, the cosine squared between the reference vectors in $\mathcal{PT}$-symmetric subspace is measured by counting the postselected outputs as shown in Eqn.~\eqref{CosSquredExp}:

\begin{equation}
    \cos_{\mathcal{PT}}^2\left(\ket{\chi_1}, \ket{\chi_3}\right) = \frac{N\left(\ket{00}\right)}{N\left(\ket{00}\right) + N\left(\ket{10}\right)},
    \label{CosSquredExp}
\end{equation}
while the population of the $\mathcal{PT}$-symmetric subsystem is shown in Eqn.~\eqref{DecisivenessExp}, correspondingly:

\begin{equation}
\mathcal{D} = \frac{\bra{\psi_{\mathcal{PT}}}\psi_{\mathcal{PT}}\rangle}{\bra{\psi_{\mathcal{PT}}}\psi_{\mathcal{PT}}\rangle + \bra{\psi_{\mathcal{PT}}}\zeta^2\ket{\psi_{\mathcal{PT}}}} = \frac{N\left(\ket{00}\right) + N\left(\ket{10}\right)}{N\left(\ket{00}\right) + N\left(\ket{10}\right) + N\left(\ket{01}\right) + N\left(\ket{11}\right)}
\label{DecisivenessExp}
\end{equation}

\begin{figure}[h!]
\includegraphics[width=\textwidth]{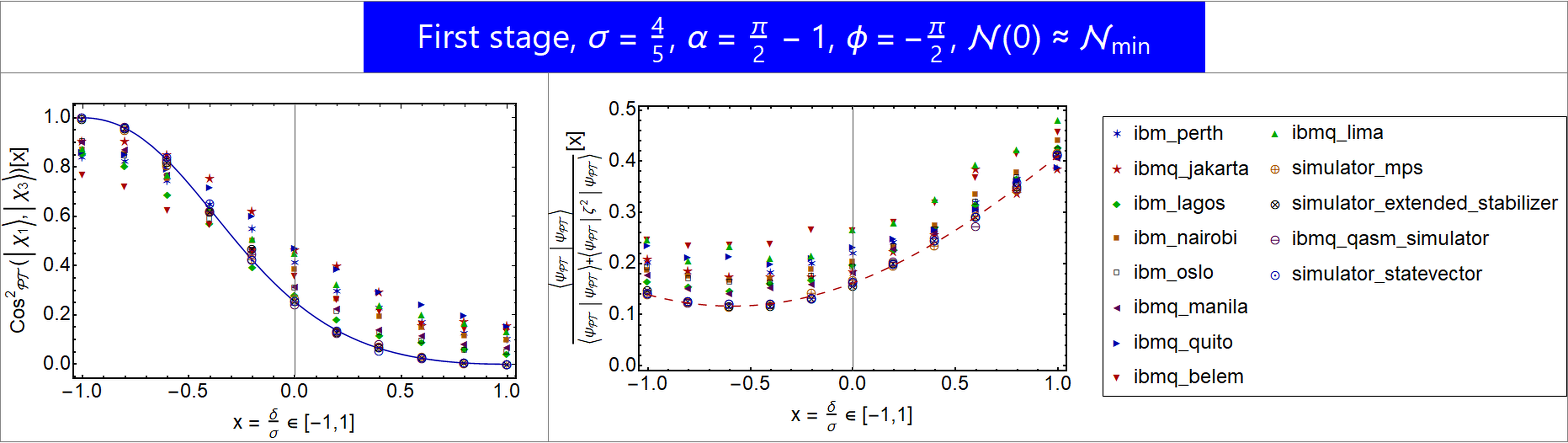}
\caption{Implementation of \textit{Stage 1} on IBM Quantum Experience using theoretical curves derived in Section~\ref{Embed}, with $\sigma = \frac{4}{5}$ corresponding to the probe state $\ket{\psi_3}$ in Eqn.~\eqref{ProbeStates}.}
\label{First45One}
\end{figure}

\begin{figure}[h!]
\includegraphics[width=\textwidth]{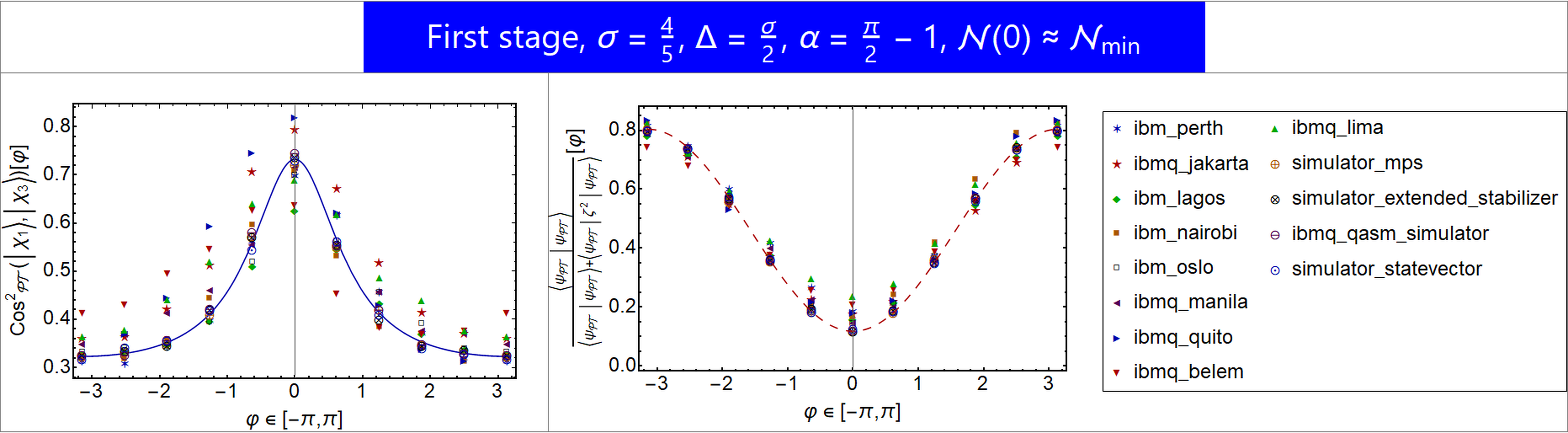}
\caption{Implementation of \textit{Stage 1} on IBM Quantum Experience using theoretical curves derived in Section~\ref{Embed}, with $\sigma = \frac{4}{5}$ and $\Delta = \frac{\sigma}{2}$ corresponding to the probe state $\ket{\psi_3^\prime}$ in Eqn.~\eqref{ProbeStates}.}
\label{First45Two}
\end{figure}

\begin{figure}[h!]
\includegraphics[width=\textwidth]{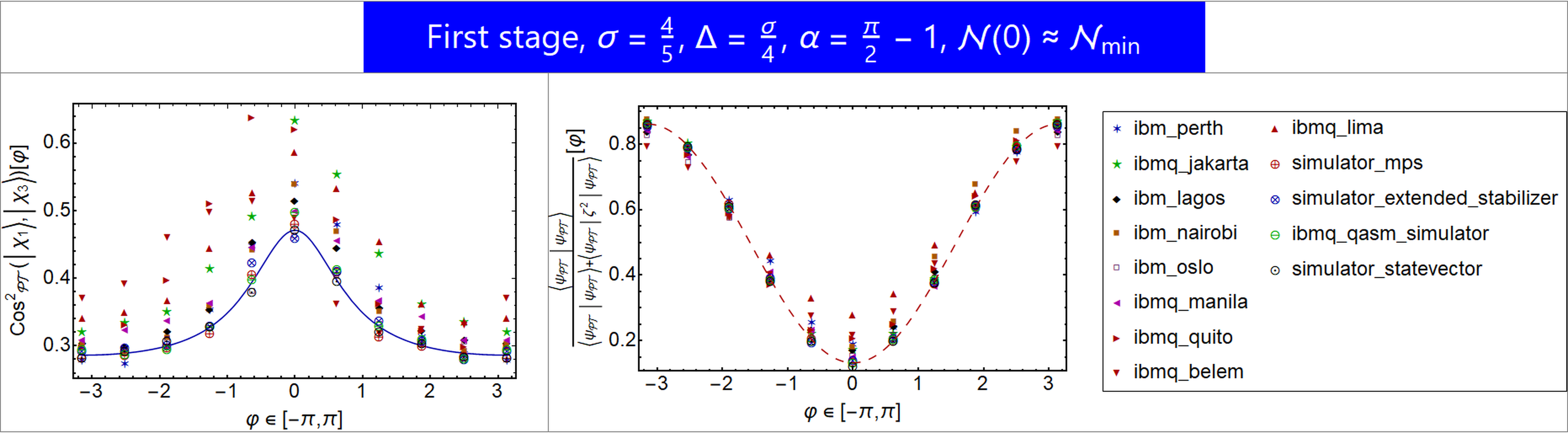}
\caption{Implementation of \textit{Stage 1} on IBM Quantum Experience using theoretical curves derived in Section~\ref{Embed}, with $\sigma = \frac{4}{5}$ and $\Delta = \frac{\sigma}{4}$ corresponding to the probe state $\ket{\psi_3^\prime}$ in Eqn.~\eqref{ProbeStates}.}
\label{First45Three}
\end{figure}

\begin{figure}[h!]
\includegraphics[width=\textwidth]{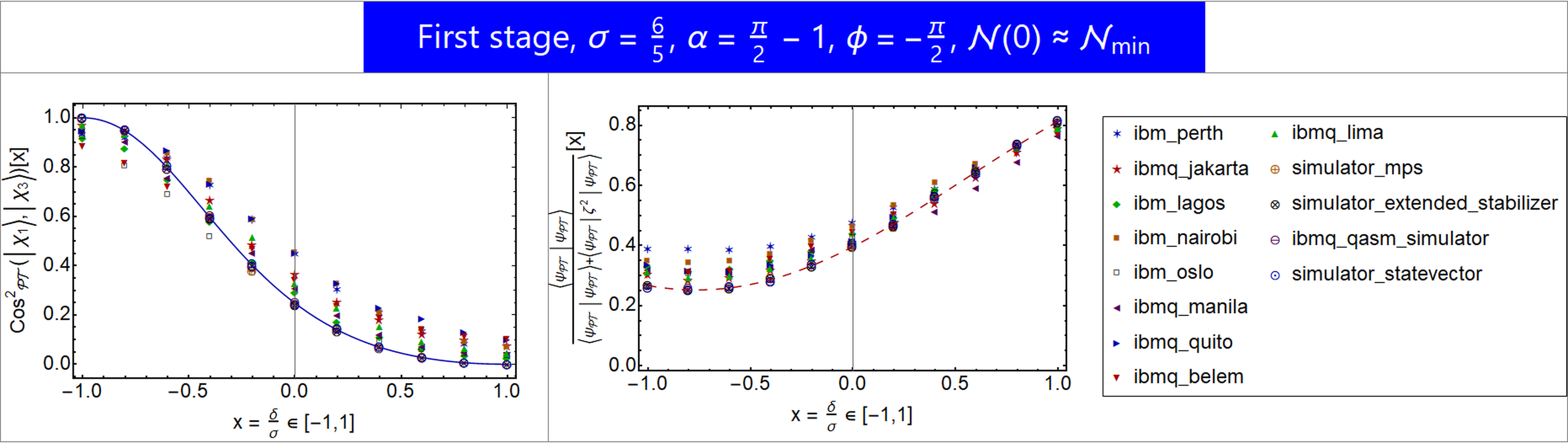}
\caption{Implementation of \textit{Stage 1} on IBM Quantum Experience using theoretical curves derived in Section~\ref{Embed}, with $\sigma = \frac{6}{5}$ corresponding to the probe state $\ket{\psi_3}$ in Eqn.~\eqref{ProbeStates}.}
\label{First65One}
\end{figure}

\begin{figure}[h!]
\includegraphics[width=\textwidth]{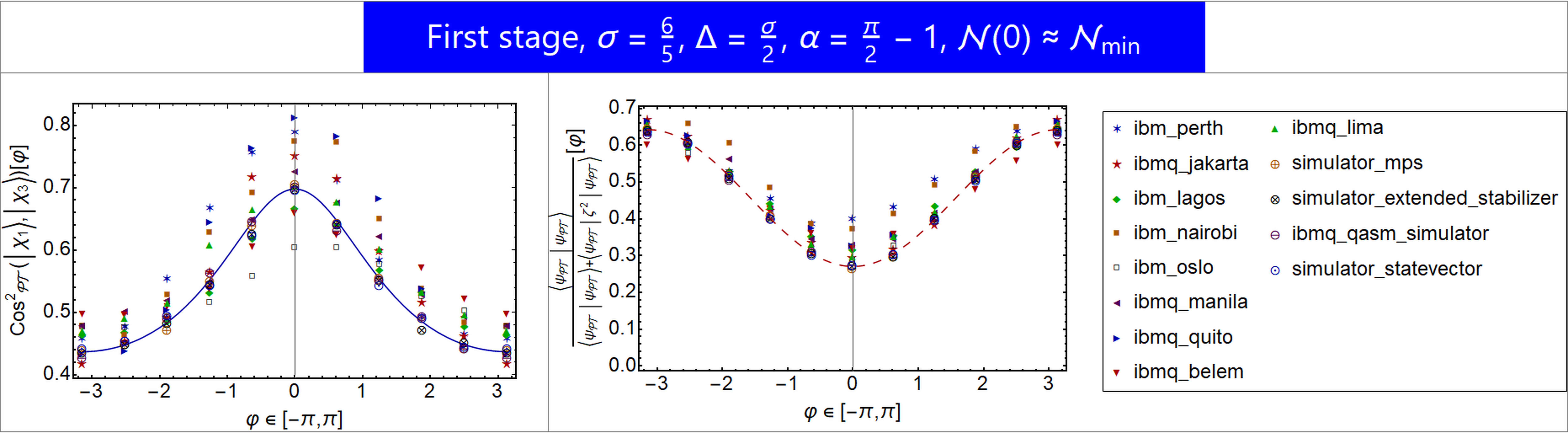}
\caption{Implementation of \textit{Stage 1} on IBM Quantum Experience using theoretical curves derived in Section~\ref{Embed}, with $\sigma = \frac{6}{5}$ and $\Delta = \frac{\sigma}{2}$ corresponding to the probe state $\ket{\psi_3^\prime}$ in Eqn.~\eqref{ProbeStates}.}
\label{First65Two}
\end{figure}

\begin{figure}[h!]
\includegraphics[width=\textwidth]{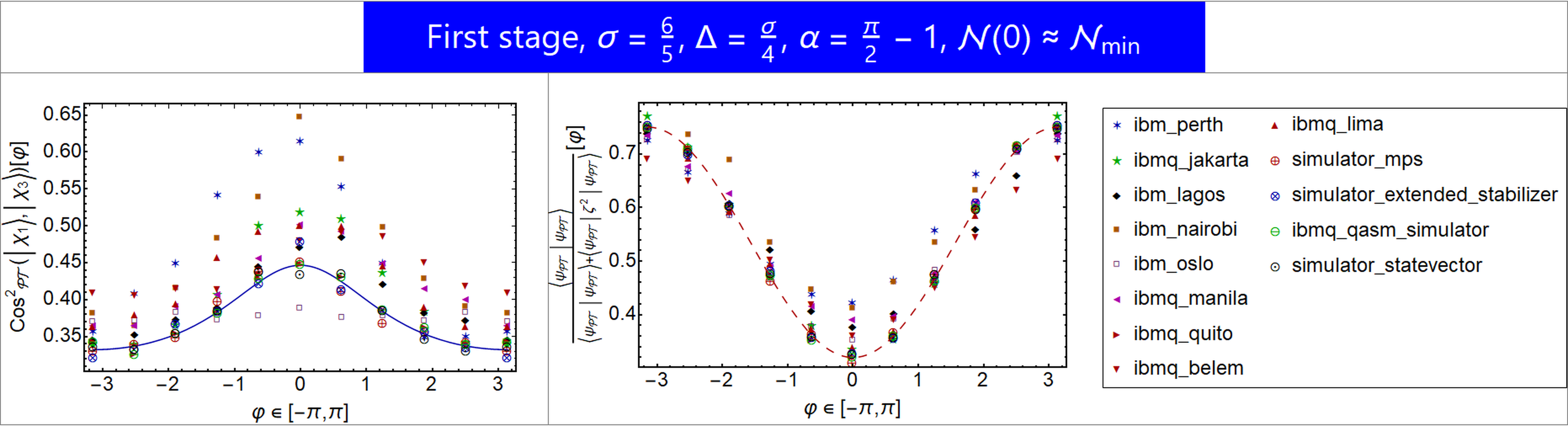}
\caption{Implementation of \textit{Stage 1} on IBM Quantum Experience using theoretical curves derived in Section~\ref{Embed}, with $\sigma = \frac{6}{5}$ and $\Delta = \frac{\sigma}{4}$ corresponding to the probe state $\ket{\psi_3^\prime}$ in Eqn.~\eqref{ProbeStates}.}
\label{First65Three}
\end{figure}

For both \textit{Stages} of our algorithm, the outputs from quantum processors are compared to the analytical results derived in Section~\ref{Embed}. We implement \textit{Stage 1} of our algorithm for $\sigma = \frac{4}{5}$ and $\frac{6}{5}$. To illustrate the changes in the geometry of the postselected space, we examine the third state in two forms, $\ket{\psi_3}$ and $\ket{\psi_3^\prime}$:
\begin{equation}
\ket{\psi_{3}}= 
\begin{pmatrix}
\cos\left(\frac{\pi + 2\delta}{4}\right)\\
 -i \sin\left(\frac{\pi + 2\delta}{4}\right)
\end{pmatrix},    \ket{\psi_{3}^\prime} =
\begin{pmatrix}
\cos\left(\frac{\pi + 2\Delta}{4}\right)\\
 e^{i\varphi} \sin\left(\frac{\pi + 2\Delta}{4}\right)
\end{pmatrix}
\label{ProbeStates}
\end{equation}
Theoretical predictions in Figs.~\ref{FirstStage1} and ~\ref{FirstStage2} corresponding to \textit{Stage 1}, using analytical results from Section~\ref{Embed}, are compared to the experimental results in Figs.~\ref{First45One},~\ref{First45Two},~\ref{First45Three},~\ref{First65One},~\ref{First65Two}, and~\ref{First65Three}.

\begin{figure}[h!]
\includegraphics[width=\textwidth]{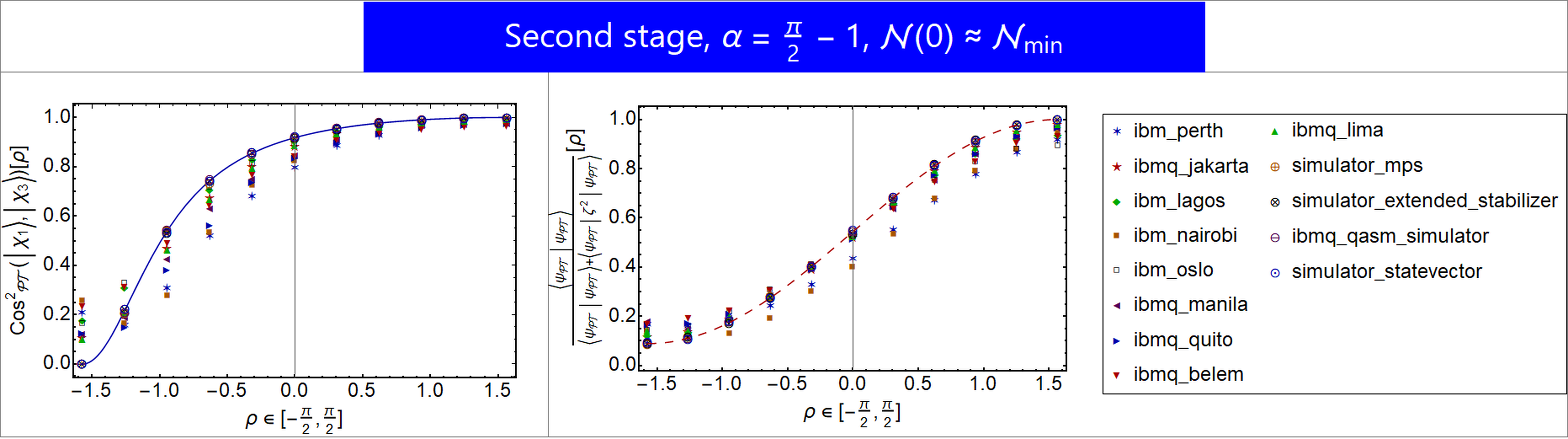}
\caption{Implementation of \textit{Stage 2} on IBM Quantum Experience with theoretical curves derived in Section~\ref{Embed}, with $\alpha = \frac{\pi}{2} - 1$
corresponding to the probe state $\ket{\chi_3}$ in Eqn.~\eqref{IntermediateAfter12}.}
\label{M10Exp}
\end{figure}

\begin{figure}[h!]
\includegraphics[width=\textwidth]{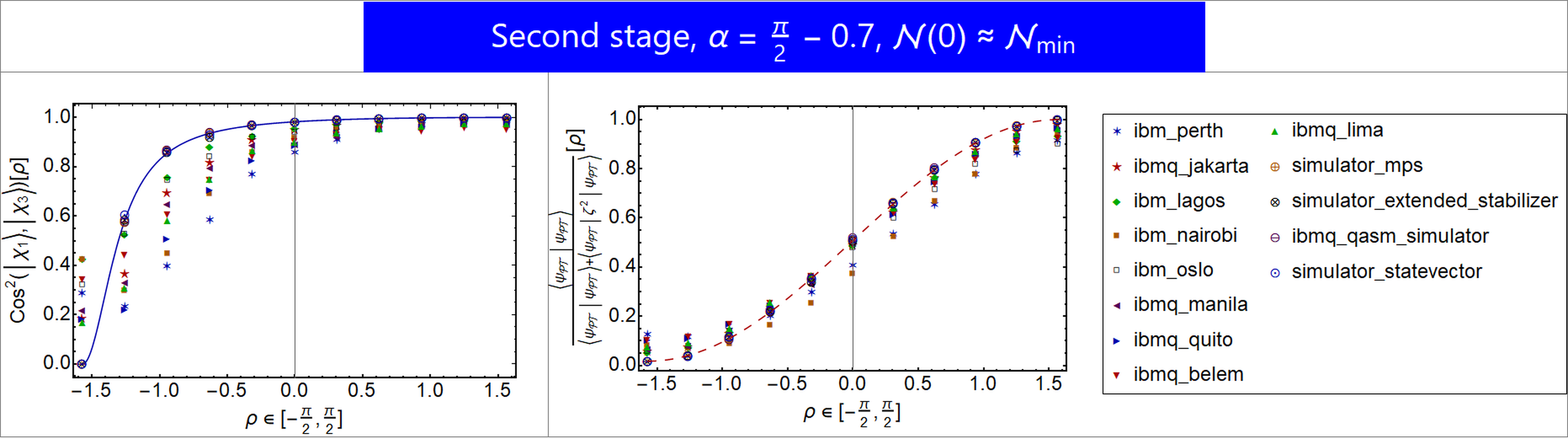}
\caption{Implementation of \textit{Stage 2} on IBM Quantum Experience with theoretical curves derived in Section~\ref{Embed}, with $\alpha = \frac{\pi}{2} - 0.7$
corresponding to the probe state $\ket{\chi_3}$ in Eqn.~\eqref{IntermediateAfter12}.}
\label{M07Exp}
\end{figure}

\begin{figure}[h!]
\includegraphics[width=\textwidth]{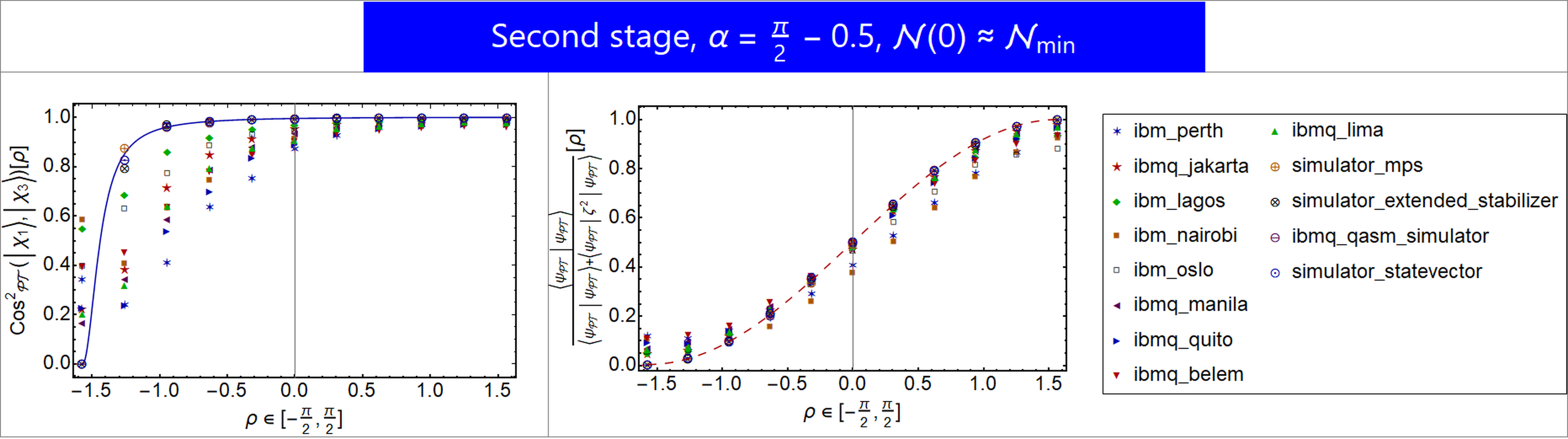}
\caption{Implementation of \textit{Stage 2} on IBM Quantum Experience with theoretical curves derived in Section~\ref{Embed}, with $\alpha = \frac{\pi}{2} - 0.5$
corresponding to the probe state $\ket{\chi_3}$ in Eqn.~\eqref{IntermediateAfter12}.}
\label{M05Exp}
\end{figure}

Similarly, we implemented \textit{Stage 2} of our algorithm with the input state given in Eqn.~\eqref{IntermediateAfter12} for $\alpha = \frac{\pi}{2} - 1$, $\alpha = \frac{\pi}{2} - 0.7$, and $\alpha = \frac{\pi}{2} - 0.5$. Fig.~\ref{SecondStageTh} shows theoretical predictions corresponding to \textit{Stage 2} as the $\alpha$ parameter approaches the exceptional point $\frac{\pi}{2}$. It can be observed that as the value of $\cos^2_{\mathcal{PT}}$ flattens out when $\alpha \rightarrow \frac{\pi}{2}$, the probability of a decisive outcome at $\rho = -\frac{\pi}{2}$ approaches zero. For \textit{Stage 2}, the corresponding comparisons between theoretical predictions and experimental results are provided in Figs.~\ref{M10Exp},~\ref{M07Exp}, and~\ref{M05Exp}.

For both \textit{Stages}, it is evident that the simulator\_mps, simulator\_extended\_stabilizer, ibmq\_qasm\_simulator, and simulator\_statevector provided by IBM processors, as indicated by circles, consistently exhibit superior agreement with theoretical predictions. Other processors, while successfully capturing the overall shapes of the theoretical curves, have significant deviations. This may imply that quantum processors employing a Matrix Product State representation, ranked-stabilizer decomposition, Open Quantum Assembly Language, and those characterizing the quantum state of a system through a state vector are particularly well-suited for performing $\mathcal{PT}$-symmetric algorithms on IBM Quantum Experience~\cite{kandala2019error}. 

\begin{figure}[h]
\includegraphics[width=\textwidth]{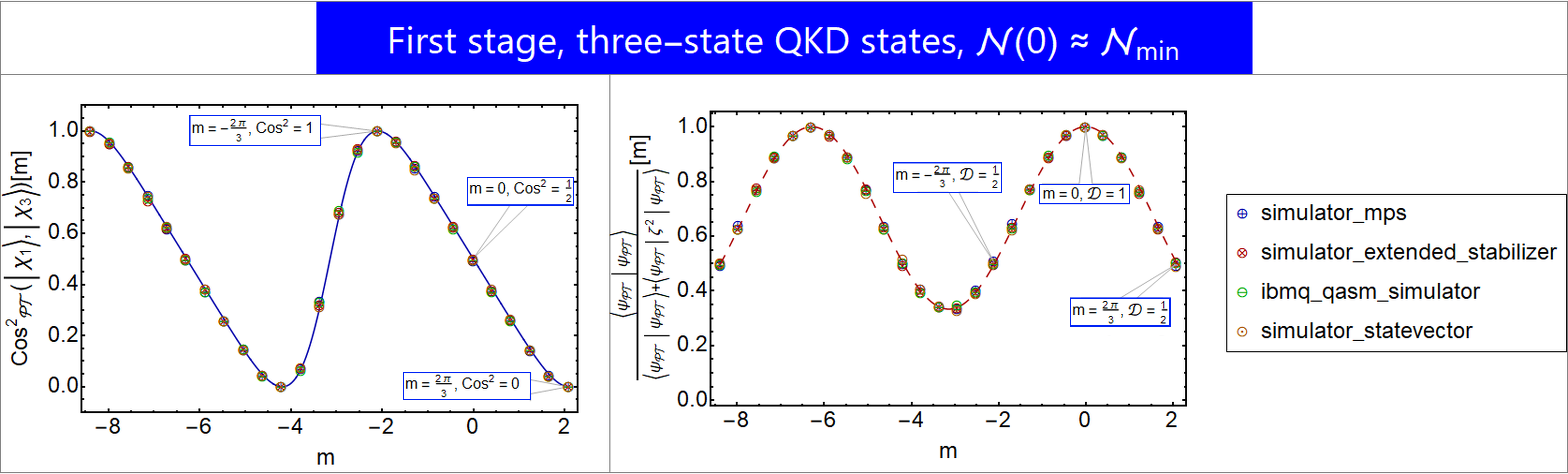}
\caption{\textit{Stage 1} for the states in Eqn.~\eqref{TwoProbeState} used in three-state QKD protocol, with $\sigma = \frac{2\pi}{3}$ and $\mathcal{N}\left(0\right)\approx\mathcal{N}_{min}$. The state corresponding to $m = 0$ always yields a conclusive result $\mathcal{D} = 1$.}
\label{ThreeQKDFig}
\end{figure}

Therefore, we used these four simulators, simulator\_mps, simulator\_extended\_stabilizer, ibmq\_qasm\_simulator, and simulator\_statevector, to perform the simulation of an attack on the three-state QKD protocol~\cite{phoenix2000three}. The states used in this QKD protocol have $\frac{2\pi}{3}$ angular separation:

\begin{equation}
|A\rangle = 
\begin{pmatrix}
1\\
0
\end{pmatrix},|B\rangle = 
\begin{pmatrix}
\frac{1}{2}\\
-\frac{\sqrt{3}}{2}
\end{pmatrix},|C\rangle = 
\begin{pmatrix}
-\frac{1}{2}\\
-\frac{\sqrt{3}}{2}
\end{pmatrix}
\label{ThreeQKDRefStates}
\end{equation}
The following gate:
\begin{equation}
K = \left(
\begin{array}{cc}
 \frac{1}{\sqrt{2}} & \frac{i}{\sqrt{2}} \\
 \frac{1}{\sqrt{2}} & -\frac{i}{\sqrt{2}} \\
\end{array}
\right),
\end{equation}
converts the reference states in Eqn.~\eqref{ThreeQKDRefStates} to our conventions in Eqn.~\eqref{TwoProbeState} as: $\ket{\psi_m, m = 0}\rightarrow\ket{A}$, $\ket{\psi_m, m = \frac{2\pi}{3}}\rightarrow\ket{B}$, $\ket{\psi_m, m = - \frac{2\pi}{3}}\rightarrow\ket{C}$. Since $\cos\left(\frac{2\pi}{3}\right) < 0$, we use Eqn.~\eqref{ThreeQKDDec} for the probability of decisive outcome.

In Fig.~\ref{ThreeQKDFig}, one can observe a remarkable agreement between theoretical predictions for \textit{Stage 1} from Section~\ref{Embed} and experimental results from simulator\_mps, 
  simulator\_extended\_stabilizer, ibmq\_qasm\_simulator, and simulator\_statevector.
For the specific geometry of states described in Eqn.~\eqref{ThreeQKDRefStates}, it can be observed that \textit{Stage 1}, with the value of $\alpha$ determined by Eqn.~\eqref{Critical} and $\mathcal{N}\left(0\right) = \mathcal{N}_{min}$, transforms these states into a mirror-symmetric configuration. In this configuration, $\ket{B}$ and $\ket{C}$ become orthogonal to each other, with $\ket{A}$ positioned between them. Regarding the probability of a decisive outcome, one can observe in Fig.~\ref{ThreeQKDFig} that $\mathcal{D}\left(\ket{B}\right) = \mathcal{D}\left(\ket{C}\right) = \frac{1}{2}$ while $\mathcal{D}\left(\ket{A}\right) = 1$. Thus, the states $\ket{B}$ and $\ket{C}$ are $50\%$ conclusive while the state $\ket{A}$ is always conclusive. In the upcoming Section~\ref{TrineQKD}, we delve into the implications of these observations and conduct a comparative analysis of our protocol against existing approaches for attacking this QKD protocol.

\section{Attack on the trine-state QKD protocol}\label{TrineQKD}

The available strategies for the attack on three-state QKD protocol are minimum error and maximum mutual information approaches~\cite{phoenix2000three}. For the geometry of states in Eqn.~\eqref{ThreeQKDRefStates}, minimum error and maximum confidence strategies coincide~\cite{barnett2009quantum}, and thus we do not consider the latter.

If the encoded state is $\ket{A}$, minimum error discrimination strategy yields correct result with the probability $\frac{2}{3}$, and misclassifies $\ket{A}$ as being $\ket{B}$ or $\ket{C}$  with the probability $\frac{1}{6}$. The same applies for $\ket{B}$ and $\ket{C}$ through the permutation $A\rightarrow B \rightarrow C$. For minimum error strategy obtaining $\ket{A}$ after the measurement, the resulting density matrix  is~\cite{phoenix2000three}:

\begin{equation}
\rho_{Min. err.} = \frac{2}{3}\ket{A}\bra{A} + \frac{1}{6}\ket{B}\bra{B} + \frac{1}{6}\ket{C}\bra{C}
\end{equation}

The maximum mutual information strategy excludes one of the states with certainty, but the other two states remain equiprobable each with $50\%$ probability and the resulting density matrix:

\begin{equation}
\rho_{Max. mut. inf.} = \frac{1}{2}\ket{\bar{B}}\bra{\bar{B}} +  \frac{1}{2}\ket{\bar{C}}\bra{\bar{C}},
\end{equation}
where $\ket{\bar{B}}$ and $\ket{\bar{C}}$ are complementary to $\ket{B}$ and $\ket{C}$~\cite{phoenix2000three}. However, both of these strategies yield the same error rate, as demonstrated by~\cite{phoenix2000three}, attributed to the inherent geometric properties of these states since:

\begin{equation}
\label{SameError}
\rho_{Min. err.} = \rho_{Max. mut. inf.} = \frac{1}{2}\ket{A}\bra{A} + \frac{1}{4}
\end{equation}

Now, consider the case when the attacker uses our $\mathcal{PT}$-symmetric approach for $N = 3$ states we developed in the previous Sections. If \textit{Stage 1}, as described in the preceding Section~\ref{Experimental}, produces an inconclusive result, the attacker immediately eliminates one of the states with $100\%$ confidence. This is because the probability of obtaining a decisive outcome for one state is $100\%$, while the other two states are equiprobable, as illustrated in Fig.~\ref{ThreeQKDFig}. For a particular choice in Fig.~\ref{ThreeQKDFig}, the attacker eliminates the state $\ket{A}$ leaving $\ket{B}$ and $\ket{C}$ equiprobable. The probability of an inconclusive result in \textit{Stage 1} is given by:

\begin{equation}
    p\left(\ket{1}^{\rm{I}}_{ancilla}\right) = \frac{1}{3}\cdot\frac{1}{2} + \frac{1}{3}\cdot0 + \frac{1}{3}\cdot\frac{1}{2} = \frac{1}{3}
\end{equation}
Thus, with probability $\frac{1}{3}$, our approach yields the result equivalent to the maximum mutual information strategy.

In case the first postselection is successful, which happens with $\frac{2}{3}$ probability, the resulting postselected geometry of the states is mirror-symmetric. However, since the postselection probability is nonuniform and varies for different states, their prior probabilities used as input for the next \textit{Stage} change from equiprobable to the values:

\begin{equation}
    p\left(\ket{A}, \ket{0}^{\rm{I}}_{ancilla}\right) = \frac{1\cdot \frac{1}{3}}{\frac{2}{3}} = \frac{1}{2}, p\left(\ket{B}, \ \ket{0}^{\rm{I}}_{ancilla}\right) = p\left(\ket{C}, \ket{0}^{\rm{I}}_{ancilla}\right) = \frac{\frac{1}{2}\cdot \frac{1}{3}}{\frac{2}{3}} = \frac{1}{4}
    \label{Post1}
\end{equation}
At this point, the attacker may choose to apply the strategy for mirror-symmetric configuration~\cite{andersson2002minimum} with $p = \frac{1}{4}$. In this scenario, the success probability rate remains $\frac{2}{3}$, consistent with the original attack outlined in~\cite{phoenix2000three}, as indicated by Eqn.(14) in~\cite{andersson2002minimum}. Despite the change in the geometry of the postselected space, the success rate remains unchanged due to the varying probabilities of successful postselections for the states in Eqn.~\eqref{ThreeQKDRefStates}.

Alternatively, if the attacker proceeds with \textit{Stage 2}, as discussed in Section~\ref{Embed}, one of the states -- $\ket{B}$ or $\ket{C}$ -- will have a $100\%$ probability of successful postselection, depending on whether $\alpha$ is greater or less than zero. Thus, if postselection of the \textit{Stage 2} fails, the attacker immediately eliminates one of these states. For definiteness, let $\alpha > 0$ and consider the probability of a decisive outcome as given in Eqn.~\eqref{Dec1}. In this case, one finds:

\begin{equation}
1 - \mathcal{D}^{\rm{II}}_+\left(\alpha,\rho = -\frac{\pi}{2}\right) = 2\left(1 - \mathcal{D}^{\rm{II}}_+\left(\alpha,\rho = 0\right)\right) = \frac{4 \sin (\alpha )}{(1 + \sin (\alpha ))^2}
\label{GeoPostF2}
\end{equation}
Considering Eqns.~\eqref{Post1} and~\eqref{GeoPostF2}, it is observed that the states $\ket{A}$ and $\ket{C}$ become equiprobable while the state $\ket{B}$ is eliminated:
\begin{equation}
p\left(\ket{A}, \ket{1}^{\rm{II}}_{ancilla}\right)p\left(\ket{A}, \ket{0}^{\rm{I}}_{ancilla}\right) = p\left(\ket{C}, \ket{1}^{\rm{II}}_{ancilla}\right)p\left(\ket{C}, \ket{0}^{\rm{I}}_{ancilla}\right)
\end{equation}
As a result, the scenario in which the second postselection fails, $\ket{1}^{\rm{II}}_{ancilla}$, is equivalent to a maximum mutual information strategy.

If the postselection for \textit{Stage 2} is successful but the measurement returns the value corresponding to the projection on $\rho = \frac{\pi}{2}$, the attacker eliminates the state $\ket{C}$ corresponding to $\rho = -\frac{\pi}{2}$ since in this case $\cos^2_{\mathcal{PT}}\left(\rho = -\frac{\pi}{2}\right) = 0$, as illustrated in Fig.~\ref{SecondStageTh}. Similarly, the remaining two states, $\ket{A}$ and $\ket{B}$, remain equiprobable since:

\begin{equation}
p\left(\ket{B}, \ket{0}^{\rm{I}}_{ancilla}\right)\cos_{\mathcal{PT}}^2\left(\kappa_{13}, \rho = \frac{\pi}{2}\right)\mathcal{D}^{\rm{II}}_+\left(\alpha,\rho = \frac{\pi}{2}\right) = \frac{1}{4}, 
\end{equation}
\begin{equation}
p\left(\ket{A}, \ket{0}^{\rm{I}}_{ancilla}\right)\cos_{\mathcal{PT}}^2\left(\kappa_{13}, \rho = 0\right)\mathcal{D}^{\rm{II}}_+\left(\alpha,\rho = 0\right) = \frac{1}{4}
\end{equation}
Finally, if the postselection at the  \textit{Stage 2} is successful, and the measurement yields the state with $\rho = -\frac{\pi}{2}$ projection corresponding to $\ket{C}$, the state $\ket{B}$ is excluded. Similarly, one observes that:
\begin{equation}
p\left(\ket{C}, \ket{0}^{\rm{I}}_{ancilla}\right)\mathcal{D}^{\rm{II}}_+\left(\alpha,\rho = -\frac{\pi}{2}\right)  =  p\left(\ket{A}, \ket{0}^{\rm{I}}_{ancilla}\right)\mathcal{D}^{\rm{II}}_+\left(\alpha,\rho = 0\right)\cos_{\mathcal{PT}}^2\left(\kappa_{23}, \rho = 0\right),
\end{equation}
and thus the states $\ket{A}$ and $\ket{C}$ remain equiprobable again.

In summary, coupling \textit{Stage 1} of our algorithm with a strategy for discriminating mirror-symmetric states, as developed in~\cite{andersson2002minimum}, results in an outcome equivalent to the maximum mutual information strategy in $\frac{1}{3}$ of the cases and, in $\frac{2}{3}$ of the cases, yields the same result as the minimum-error strategy. If both \textit{Stages} are employed, our approach yields an equivalent result to the maximum mutual information strategy. Given that the minimum error and maximum mutual information strategies exhibit the same error rate for this QKD protocol due to Eqn.~\eqref{SameError}, our algorithm achieves precisely the same error rate as these strategies in both cases. 

However, despite having the same error rate as conventional Hermitian approaches, as discussed in~\cite{wang2024demonstration}, $\mathcal{PT}$-symmetric quantum state discrimination is advantageous in terms of the quantum resources involved. Additionally, while our algorithm does not provide an advantage for the specific states used in the three-states QKD protocol, as given in Eqn.~\eqref{ThreeQKDRefStates}, it can be advantageous in other scenarios. In cases involving highly nonsymmetric states, where explicit solutions are not readily available and intricate computations are required~\cite{samsonov2009minimum}, our approach's ability to map three arbitrary states to a predefined and standardized set can be beneficial in practical applications.

In the following Sections, using the results derived earlier, we identify applications where our algorithm offers significant \textit{technical} advantages over its Hermitian counterparts.

\section{Applications for postselected $\mathcal{PT}$-symmetric sensing}\label{SensingApplication}

\begin{figure}[h]
\includegraphics[width=\textwidth]{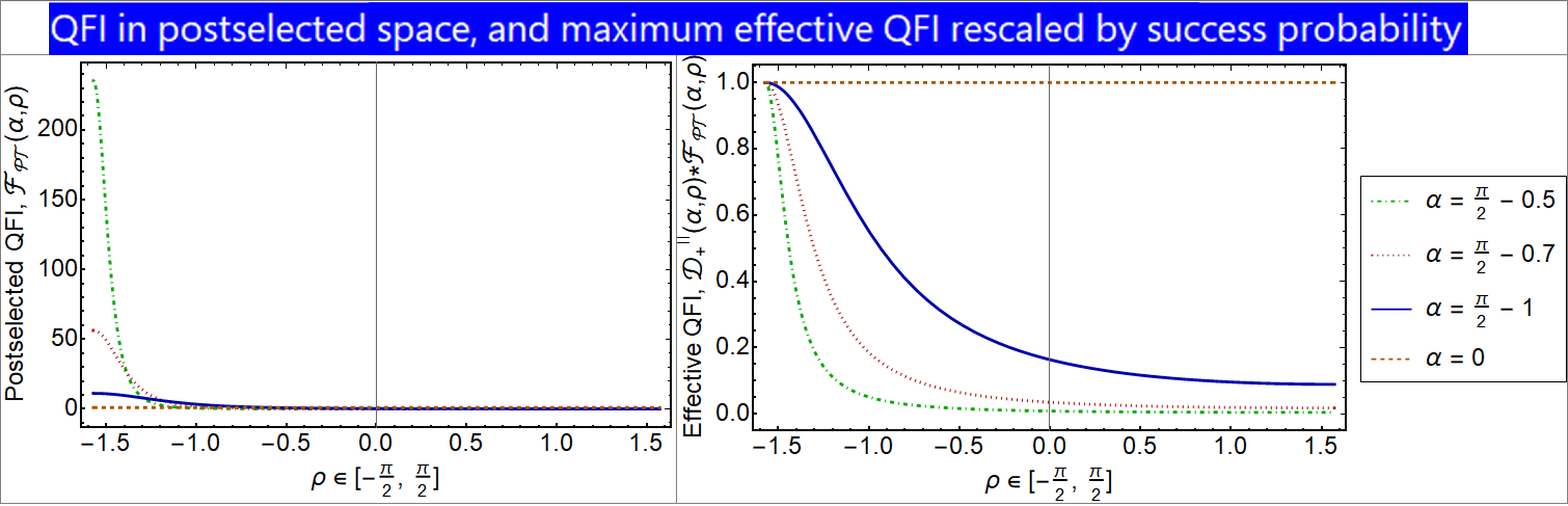}
\caption{Quantum Fisher information upon successful postselection $\mathcal{F}_{\mathcal{PT}}\left(\alpha,\rho\right)$, and its effective value rescaled by probability of successful postselection $\mathcal{D}^{\rm{II}}_+\left(\alpha,\rho\right)$. }
\label{QFI_and_eff_QFI}
\end{figure}

With the help of explicit expressions for the probability of a decisive outcome in $\mathcal{PT}$-symmetric evolution derived in Section~\ref{Embed}, we can now assess its implications for quantum sensing applications. The use of non-Hermitian single-qubit gates for quantum sensing has been proposed~\cite{chu2020quantum} because such systems exhibit divergent susceptibility, which promises increased sensitivity. However, probabilistic metrology, evaluated by the mean square error of estimation (MSE) and QFI, cannot surpass the quantum limits of single-parameter estimation, as shown in~\cite{combes2014quantum}. Recent work has confirmed that when postselection probability is taken into account, the \textit{average} QFI and susceptibility do not increase~\cite{ding2023fundamental}.  As noted in~\cite{ding2023fundamental}, postselected metrology does not outperform Hermitian metrology when resources are unlimited. Although postselected metrology can potentially outperform Hermitian metrology in practical scenarios where resources are limited~\cite{ding2023fundamental}, the exact conditions for this were not specified in~\cite{ding2023fundamental}. This is the subject of our paper and is specified further in the text.

In the conventional Hermitian approach, when measuring the value of $\rho\in\left[-\frac{\pi}{2}, \frac{\pi}{2}\right]$, two projectors on two orthogonal directions saturate the Cramer-Rao inequality~\cite{chapeau2016optimizing}. We supplement these projectors with the $\mathcal{PT}$-symmetric transformation to explore possible benefits it can provide. Our results are consistent with the findings of all the aforementioned works. Namely, in Fig.~\ref{SecondStageTh}, one can observe a sharp spike in sensitivity at $\rho = -\frac{\pi}{2}$ in terms of $\cos^2_{\mathcal{PT}}$, which resembles ~\cite{chu2020quantum}. Similarly to~\cite{guo2017enhancing,wang2020quantum}, for the density matrix $\rho_{init}$ corresponding to the pure state $\ket{\chi_3}$ in Eqn.~\eqref{IntermediateAfter12}, we compute QFI, $\mathcal{F}_\rho^{Pure}$, after the \textit{Stage 2} of $\mathcal{PT}$-symmetric evolution:
\begin{equation}
\rho^{Stage \ 2}\left(t\right) =  \frac{e^{-i\mathcal{H}t}\rho_{init}e^{i\mathcal{H}^{\dagger}t}}{Tr\left(e^{-i\mathcal{H}t}\rho_{init}e^{i\mathcal{H}^{\dagger}t}\right)}, \ \mathcal{F}_\rho^{Pure} = 2Tr\left[\left(\partial_\rho \rho^{Stage \ 2}\left(t = \tau^{\rm{II}} = \frac{\pi}{2\omega}\right)\right)^2\right]
\end{equation}
One observes at $\rho = -\frac{\pi}{2}$ an apparently divergent QFI, similarly to~\cite{chu2020quantum}:
\begin{equation}
\mathcal{F}_{\mathcal{PT}}\left(\alpha,\rho\right) = \frac{4 \cos ^4(\alpha )}{(3 + 4 \sin (\alpha ) \sin (\rho )-\cos (2 \alpha ))^2} = \frac{\left(\alpha -\frac{\pi }{2}\right)^4}{4 (1 + \sin (\rho ))^2}+O\left(\left(\alpha -\frac{\pi }{2}\right)^5\right),
\end{equation}
However, by an explicit computation, one obtains:

\begin{equation}
\mathcal{D}^{\rm{II}}_+\left(\alpha,\rho\right)\mathcal{F}_{\mathcal{PT}}\left(\alpha,\rho\right) = \frac{4 \cos ^4(\alpha )}{(3 + 4 \sin (\alpha ) - \cos (2 \alpha )) (3 + 4 \sin (\alpha ) \sin (\rho ) - \cos (2
   \alpha ))}
\end{equation}
As shown in Fig.~\ref{QFI_and_eff_QFI}, the QFI in $\mathcal{PT}$-symmetric postselected space, $\mathcal{F}_{\mathcal{PT}}\left(\alpha,\rho\right)$, depicted on the left, is significantly higher than that in the Hermitian case corresponding to $\alpha = 0$. However, at $\rho = -\frac{\pi}{2}$, the \textit{average} QFI rescaled by the probability of successful postselection, remains the same as in the Hermitian case. Additionally, as observed, when $\rho > -\frac{\pi}{2}$, the effective QFI is actually smaller than that in the Hermitian case. Therefore, the application of the $\mathcal{PT}$-symmetric exceptional point does not provide an advantage over the Hermitian case in terms of \textit{average} QFI, in complete agreement with~\cite{combes2014quantum} and~\cite{ding2023fundamental}. However, as discussed in~\cite{jordan2014technical}, even though the average QFI after postselection remains the same as in the Hermitian case, the ability to ``condense'' all QFI about the detected parameter into a small fraction of events provides significant \textit{technical} advantages, and we quantify these advantages further in the text.

As discussed in~\cite{arvidsson2020quantum}, any real experiment involves the preparation cost $\left(\mathcal{C}_P\right)$, the final measurement cost $\left(\mathcal{C}_M\right)$, and, in the case of postselected metrology, the postselection cost $\left(\mathcal{C}_{ps}\right)$. Thus, a more reasonable figure of merit representing experimental realities is the information-cost rate, $R^{\text{Herm.}}$ for the Hermitian system, and $R^{\mathcal{PT}}$ for the $\mathcal{PT}$-symmetric system, respectively:

\begin{equation}
    R^{\text{Herm.}} = \frac{\mathcal{I}}{\mathcal{C}_P + \mathcal{C}_M}, \quad 
    R^{\mathcal{PT}} = \frac{p^{\text{ps}} \mathcal{I}^{\text{ps}}}{\mathcal{C}_P + \mathcal{C}_{\text{ps}} + p^{\text{ps}} \mathcal{C}_M}, \label{Rates}
\end{equation}
where $\mathcal{I} = 1$ is the QFI in the Hermitian system, $\mathcal{I}^{ps}$ is the QFI upon successful postselection in the $\mathcal{PT}$-symmetric system, $\mathcal{I}^{ps} = \mathcal{F}_{\mathcal{PT}}(\alpha, \rho)$, and $p^{ps} = \mathcal{D}^{\rm{II}}_+(\alpha, \rho)$ is the probability of successful postselection, as defined in our paper.

In the case when $\mathcal{C}_M \gg \mathcal{C}_P + \mathcal{C}_{ps}$, as one observes from Eqn.~\eqref{Rates} and Fig.~\ref{QFI_and_eff_QFI}: 

\begin{equation}
   R^{\mathcal{PT}} 
   \underbrace{\longrightarrow}_{\mathcal{C}_M \gg \mathcal{C}_P + \mathcal{C}_{ps}} 
   \frac{\mathcal{F}_{\mathcal{PT}}(\alpha, \rho)}{\mathcal{C}_M} 
   \gg R^{\text{Herm.}} \approx \frac{\mathcal{I}}{\mathcal{C}_M},
   \quad \text{for } \alpha \rightarrow \frac{\pi}{2}, \; \rho \approx -\frac{\pi}{2}.
\end{equation}
and the information-cost rate of $\mathcal{PT}$-symmetric system drastically outperforms its Hermitian counterpart, similarly to the conclusions of~\cite{arvidsson2020quantum}. 

\begin{figure}[h]
\includegraphics[width=\textwidth]{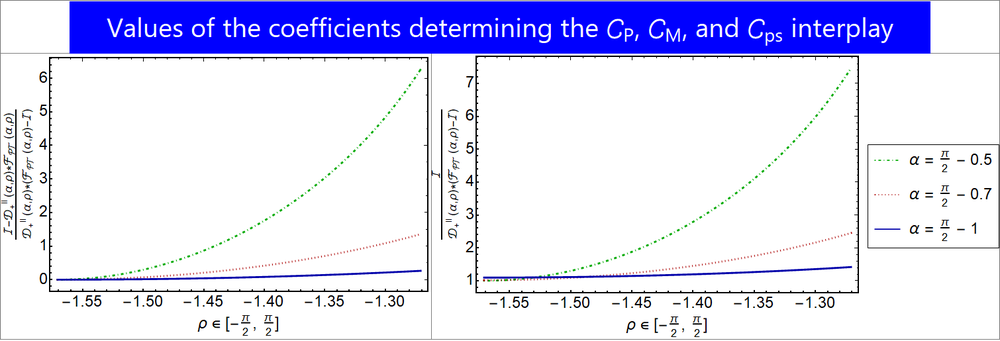}
\caption{Illustrative values of the coefficients in Eqn.~\eqref{CMCPCpsCoeff} in the vicinity of the exceptional point.}
\label{CMCPCps}
\end{figure}

However, other results from~\cite{arvidsson2020quantum} aimed at relaxing the condition $\mathcal{C}_M \gg \mathcal{C}_P + \mathcal{C}_{ps}$ are inapplicable to the qubit case, as the unitary transformation encoding the qubit state has exactly two eigenvalues. This condition may be too demanding in real-life experiments. Nevertheless, the condition under which a $\mathcal{PT}$-symmetric system outperforms its Hermitian counterpart in terms of the information-cost rate can be significantly relaxed by explicitly comparing $R^{Herm.}$ and $R^{\mathcal{PT}}$ in Eqn.~\eqref{Rates}:

\begin{equation}
    \mathcal{C}_M \ge \frac{\mathcal{I}  -  \mathcal{D}^{\rm{II}}_+\left(\alpha,\rho\right) \mathcal{F}_{\mathcal{PT}}\left(\alpha,\rho\right)}{\mathcal{D}^{\rm{II}}_+\left(\alpha,\rho\right) (\mathcal{F}_{\mathcal{PT}}\left(\alpha,\rho\right) - \mathcal{I})} \mathcal{C}_P + \frac{\mathcal{I}}{\mathcal{D}^{\rm{II}}_+\left(\alpha,\rho\right) (\mathcal{F}_{\mathcal{PT}}\left(\alpha,\rho\right) - \mathcal{I})}\mathcal{C}_{ps}, 
    \label{CMCPCpsCoeff}
\end{equation}
with the coefficients' explicit expressions being:  

\begin{equation}
\begin{cases}
\frac{\mathcal{I}  -  \mathcal{D}^{\rm{II}}_+\left(\alpha,\rho\right) \mathcal{F}_{\mathcal{PT}}\left(\alpha,\rho\right)}{\mathcal{D}^{\rm{II}}_+\left(\alpha,\rho\right) (\mathcal{F}_{\mathcal{PT}}\left(\alpha,\rho\right) - \mathcal{I})} = -\frac{(1 + \sin (\alpha ))^2 (1 + \sin (\rho ))}{\sin (\alpha ) (3-\cos (2 \rho ))+(3-\cos (2 \alpha ))
   \sin (\rho )}    \\
   \frac{\mathcal{I}}{\mathcal{D}^{\rm{II}}_+\left(\alpha,\rho\right) (\mathcal{F}_{\mathcal{PT}}\left(\alpha,\rho\right) - \mathcal{I})} = -\frac{\csc (\alpha ) (3 + 4 \sin (\alpha )-\cos (2 \alpha )) (3 + 4 \sin (\alpha ) \sin (\rho )-\cos (2
   \alpha ))}{16 (\sin (\alpha )+\sin (\rho )) (1 + \sin (\alpha ) \sin (\rho ))}
\end{cases} \label{CoeffExp}  
\end{equation}
As observed in Fig.~\ref{CMCPCps}, in the vicinity of the exceptional point when $\alpha \rightarrow \frac{\pi}{2}$ and $\rho \approx -\frac{\pi}{2}$, the coefficients in Eqn.~\eqref{CoeffExp} approximately equal:

\begin{equation}
\mathcal{I}  -  \mathcal{D}^{\rm{II}}_+\left(\alpha,\rho\right) \mathcal{F}_{\mathcal{PT}}\left(\alpha,\rho\right)\approx 0, \   \frac{\mathcal{I}}{\mathcal{D}^{\rm{II}}_+\left(\alpha,\rho\right) (\mathcal{F}_{\mathcal{PT}}\left(\alpha,\rho\right) - \mathcal{I})} \approx 1, 
\end{equation}
and thus, the condition $\mathcal{C}_M \gg \mathcal{C}_P + \mathcal{C}_{ps}$ is relaxed to the much weaker condition of $\mathcal{C}_M \gtrsim \mathcal{C}_{ps}$. In single-particle experiments, post-sampling can be practically free and save resources by eliminating the need to run the final measurement in case of failed post-selection~\cite{alonso2019trace}, and thus our results are directly applicable to current experimental designs.

As shown in Fig.~\ref{CMCPCps}, the corresponding coefficients become large as one departs from the exceptional point. Therefore, $\mathcal{PT}$-symmetric sensors should be used sufficiently close to the exceptional point to outperform their Hermitian counterparts. Our results in Eqns.~\eqref{CMCPCpsCoeff} and~\eqref{CoeffExp} provide experimenters with the necessary tools to evaluate the performance of an arbitrary $\mathcal{PT}$-symmetric system using three reference states for single-parameter estimation tasks.

The next Section assesses the technical merits of our approach for the punctuated quantum database search.

\section{Application for punctuated unstructured database search}\label{SearchApplication}

Finally, consider our approach in conjunction with the search over an unstructured database of size $M = 2^n$. The renowned Grover's search algorithm~\cite{grover1996proceedings} finds the solution in time $\sim \sqrt{M}$ and is optimal in terms of the number of oracle calls required for the search process~\cite{zalka1999grover}.

As the quantum state of Grover's algorithm remains in a two-dimensional subspace after each application of the oracle~\cite{grover1996proceedings}, our results developed for the qubit case in Section~\ref{Embed} are directly applicable to the unstructured database search as well. Let $\ket{\omega}$ be the state to be identified, $\ket{s} = \frac{1}{\sqrt{M}}\sum_{x = 0}^{M - 1} \ket{x}$ be the initial state consisting of all possible options before applying the oracles, and $\ket{s^\prime} = \frac{1}{\sqrt{M - 1}}\sum_{x \ne \omega} \ket{x}$ be the state orthogonal to $\ket{\omega}$. In the limit $M\gg 1$, the state of the system after applying the oracle $k$ times, in the two-dimensional basis formed by orthogonal states $\ket{s^\prime}$ and $\ket{\omega}$, is given by~\cite{grover1996proceedings}:

\begin{equation}\label{GroverState}
\ket{\Psi_G} = 
\begin{pmatrix}
 \cos\left(\frac{2k}{\sqrt{M}}\right)
 \\
\sin\left(\frac{2k}{\sqrt{M}}\right)
\end{pmatrix}
\end{equation}
After applying the oracle $k$ times, the probability of correctly identifying the state is: \begin{equation}
p_+\left(k\right) = \sin^2\left(\frac{2k}{\sqrt{M}}\right),   
\end{equation}
and approaches $100\%$ when $k = \frac{\pi}{4}\sqrt{M}$~\cite{grover1996proceedings}, and $ \ket{\Psi_G\left(k = \frac{\pi}{4}\sqrt{M}\right)} = \ket{1}$.

However, in practice, it is possible that the cycle of Grover's search algorithm is not completed, and the final measurement is performed \textit{before} the state of the system represented by $\ket{\Psi_G}$ reaches $\ket{1}$. For example, when only \textit{half} the number of oracles needed to perform the complete cycle of Grover's search algorithm is used, the state of the system is:
\begin{equation}
\ket{\Psi_G \left(k = \frac{1}{2}\times\frac{\pi}{4}\sqrt{M}\right)} = 
\frac{1}{\sqrt{2}}\begin{pmatrix}
1 \\
1
\end{pmatrix},
\end{equation}
and the applied measurement will yield the correct solution $50\%$ of the time~\cite{gingrich2000generalized, boyer1998tight}. The average number of oracle calls needed to find the solution remains the same as in the original Grover's algorithm since the final measurement must be repeated twice on average.

\begin{figure}[h]
\includegraphics[width=0.6\textwidth]{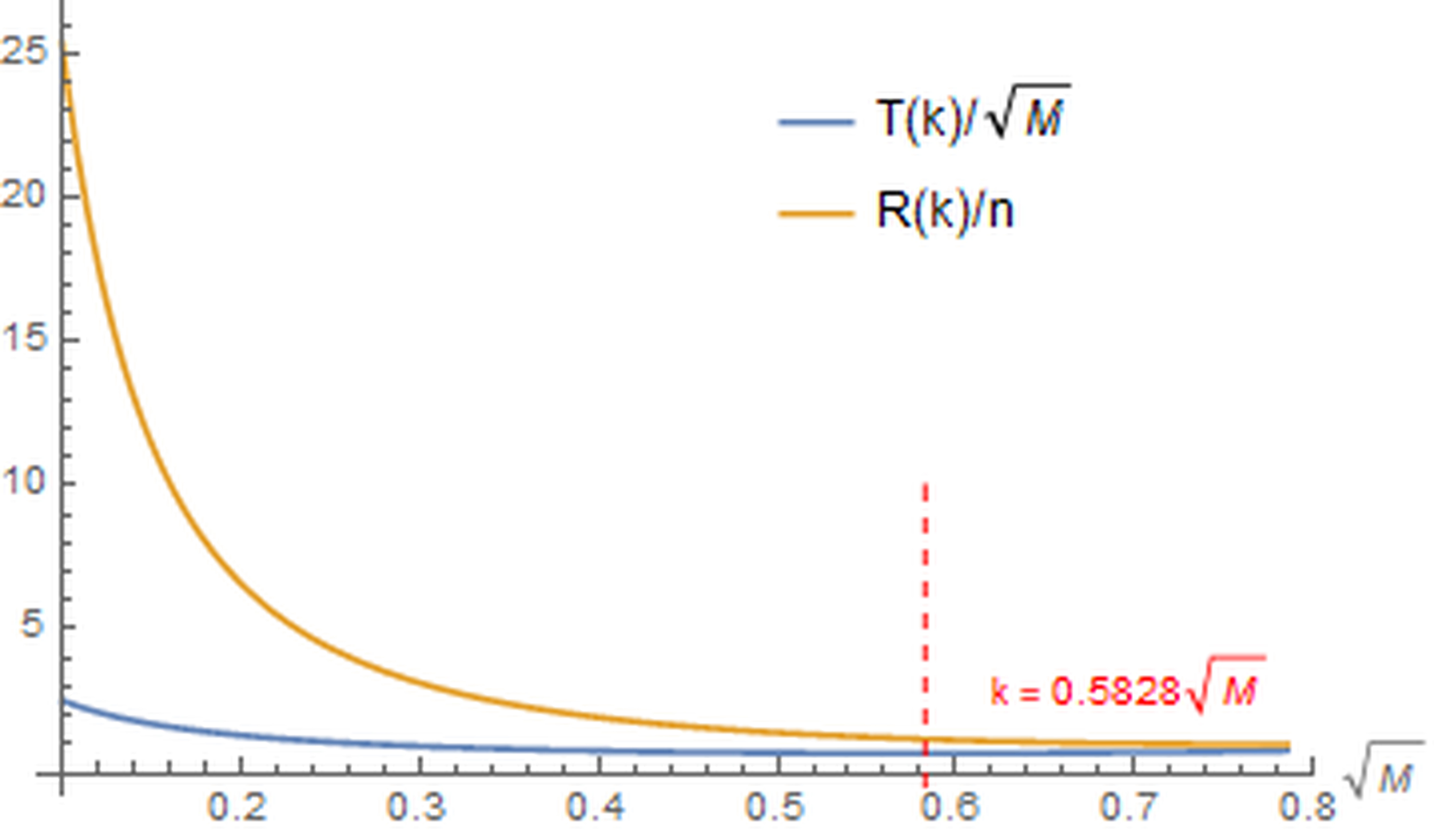}
\caption{The normalized average number of oracle calls, $\frac{T\left(k\right)}{\sqrt{M}}$, and the average number of the final measurements per qubit, $\frac{R\left(k\right)}{n}$, needed to find the solution using the punctuated version of Grover's search algorithm.}
\label{GrFig}
\end{figure}
Reducing the depth of Grover's search algorithm can provide a decrease in the \textit{average} number of oracle calls~\cite{gingrich2000generalized,boyer1998tight} by minimizing:

\begin{equation}\label{T(k)}
    T\left(k\right) = \sum_{i = 1}^\infty \left(1 - p_+\left(k\right)\right)^{i - 1}p_+\left(k\right) i k = k\csc ^2\left(\frac{2 k}{\sqrt{M}}\right)  \rightarrow min
\end{equation}
This improvement capitalizes on the observation that the convergence towards the end of the complete Grover's search algorithm is slow, as indicated in Eqn.~\eqref{GroverState}. The search is stopped after approximately $0.5828\sqrt{M}$ oracle applications, as illustrated in Fig.~\ref{GrFig}, earlier than in the original version of Grover's algorithm $\left(\frac{\pi\sqrt{M}}{4}\right)$. While the probability of correctly identifying the state is approximately $84.458\%$, with the risk of having to restart the search, the \textit{average} number of oracle calls is reduced by $12\%$~\cite{gingrich2000generalized,boyer1998tight}. 

Additionally, due to technical limitations in the depth of the quantum circuit, it may be necessary to perform the final measurement before completing the entire Grover's cycle. The depth parameter is crucial for Noisy Intermediate Scale Quantum (NISQ) computers defined by Preskill~\cite{preskill2018quantum}. These quantum computers feature noisy qubits and have the potential to solve practical problems of commercial significance faster than conventional supercomputers or with lower energy consumption. To address the challenges posed by error accumulation, decoherence, and error correction, it is recommended that the corresponding quantum circuits exhibit a shallow depth corresponding to a small number of qubit gate cycles~\cite{ezratty2023we}. Additionally, the NIST call for proposals on Post-Quantum Cryptography (PQC)~\cite{nist2016,alagic2022status} emphasizes a specific form of the quantum circuit model. In this variant, the adversary is constrained to executing a maximum of MAXDEPTH gates in series.

In addition to the number of oracle calls and the depth of the quantum circuit, an important parameter in the practical implementation of Grover's search algorithm is the number of qubit readouts. These readouts convert quantum information into classical information, allowing the states of the qubit to be classified as ``0'' or ``1''. They are among the most error-prone and slowest operations on a superconducting quantum processor. Readout errors on state-of-the-art cloud-based quantum processors can be more than $10\%$ for some qubits~\cite{ibm2022quantum}, and the readout time is greater than $300ns$. For example, the quantum processors used to demonstrate the quantum supremacy~\cite{arute2019quantum} have readout errors of $9\%$~\cite{weber2022datasheet}, and the execution time over $1\mu s$~\cite{weber2022datasheet}.

If the cycle of the quantum database search is not complete, the number of times one must perform the readout of $n\gg 1$ qubits on average is given by:

\begin{equation}
R\left(k\right) = \frac{n}{p_+\left(k\right)} = n\cdot\csc^2\left(\frac{2k}{\sqrt{M}}\right)\label{R_k}
\end{equation}
In the example discussed earlier, halving the depth of the original Grover search algorithm doubles the qubit readout cost, which can be demanding for large-scale quantum database searches. As illustrated in Fig.~\ref{GrFig}, the qubit readout cost rapidly increases when using shallow-depth quantum searches.

In the remainder of this Section, we demonstrate how the usage of $\mathcal{PT}$ symmetry can drastically reduce the qubit readout cost in quantum database searches, at the cost of one extra ancilla that is used only once and not involved during the punctuated Grover's search algorithm.

The possibility to improve the search over an unstructured database by using $\mathcal{PT}$ symmetry was initially discussed in~\cite{bender2013pt} (referring to~\cite{abrams1998nonlinear}) and further discussed in~\cite{croke2015pt}. According to~\cite{abrams1998nonlinear}, the ability to exponentially separate the qubit states close to each other implies the capability to search exponentially large databases in polynomial time. However, as discussed in~\cite{croke2015pt}, achieving such an operation is only possible with an exponentially small probability of success. Consequently, it is not feasible to search over an unstructured database using fewer oracle calls than in Grover's search algorithm by applying $\mathcal{PT}$ symmetry.

Our results in Eqns.~\eqref{Leverage3} and~\eqref{Leverage4} are in complete agreement with~\cite{croke2015pt}. The following unitary transformation:

\begin{equation}
Q = \left(
\begin{array}{cc}
 \frac{1}{\sqrt{2}} & -\frac{i}{\sqrt{2}} \\
 -\frac{i}{\sqrt{2}} & \frac{1}{\sqrt{2}} \\
\end{array}
\right)\label{Q},
\end{equation}
applied in the two-dimensional basis formed  by $\{\ket{s^\prime}, \ket{\omega}\}$, maps $\ket{s^\prime}$ to $\rho = -\frac{\pi}{2}$ and $\ket{\omega}$ to $\rho = -\frac{\pi}{2}$ in the conventions of Section~\ref{Embed}. Performing the same computations in this basis as in Section~\ref{Embed} for this case, one finds the probability of correctly identifying the solution after successful postselection:

\begin{equation}
\label{pPTG}
p_+^{\mathcal{PT}}\left(k,\alpha\right) = \frac{2 \left(1 + \sin \left(\alpha \right)\right)^2 \sin ^2\left(\frac{2k}{\sqrt{M}}\right)}{3 -\cos \left(2 \alpha \right)-4
   \sin \left(\alpha \right) \cos \left(\frac{4k}{\sqrt{M}}\right)},
\end{equation}
as well as the probability of successful postselection: 
\begin{equation} 
\label{DPTG}
\mathcal{D}_G^{\mathcal{PT}}\left(k,\alpha\right) = \frac{\sec^2\left(\alpha \right)\left(1-\sin\left(\alpha \right)\right)  \left(3 - \cos \left(2
   \alpha \right)-4 \sin \left(\alpha \right) \cos \left(\frac{4k}{\sqrt{M}}\right)\right)}{2 \left(1 + \sin
   \left(\alpha \right)\right)}
\end{equation}
As one can observe in Eqn.~\eqref{pPTG}, it is possible to exponentially increase the value of $p_+^{\mathcal{PT}}\left(k,\alpha\right)$. However, this enhancement is counterbalanced by an exponential decrease in postselection probability $\mathcal{D}_G^{\mathcal{PT}}\left(k,\alpha\right)$ in Eqn.~\eqref{DPTG}, resulting in no improvement in the average number of oracle calls since:
\begin{equation}
\label{Convert}
p_+^{\mathcal{PT}}\left(k,\alpha\right)\cdot   \mathcal{D}_G^{\mathcal{PT}}\left(k,\alpha\right) = \sin ^2\left(\frac{2k}{\sqrt{M}}\right) = p_+\left(k\right)
\end{equation}

The dilation method required to execute the $\mathcal{PT}$-symmetric transformation on $n$ qubits, as described above, requires only a \underline{single} ancilla, regardless of the dimension of the underlying system~\cite{neumark}. Therefore, compared to the conventional punctuated Grover's search algorithm which requires $n$ qubits, its $\mathcal{PT}$-enhanced version requires $n + 1$ qubits. Of these, $n$ qubits are used to perform the punctuated Grover's search algorithm, while the additional qubit is used only once to execute the non-Hermitian $\mathcal{PT}$-symmetric transformation before performing the final measurement.

In the approach we propose, one applies the $\mathcal{PT}$-symmetric transformation on $n$ working qubits with $\alpha\rightarrow\frac{\pi}{2}$ achieving $p_+^{\mathcal{PT}}\left(k,\alpha\right)\rightarrow 1$ at the cost of reduced  $\mathcal{D}_G^{\mathcal{PT}}\left(k,\alpha\right)$. In this limit, from Eqn.~\eqref{Convert}, one finds that:

\begin{equation}
\mathcal{D}_G^{\mathcal{PT}}\left(k,\alpha\rightarrow\frac{\pi}{2}\right)  \rightarrow p_+\left(k\right) 
\end{equation}

After that, similarly to Section~\ref{SensingApplication}, the final measurement of $n$ qubits is precluded if the ancilla is measured to be $\ket{1}_{ancilla}$. This approach saves resources by avoiding the need for $n$ qubit measurements, which is resource-demanding as discussed above. After $\mathcal{PT}$-symmetric transformation, in the limit $\alpha\rightarrow\frac{\pi}{2}$, the state-ancilla system is:

\begin{equation}
\ket{\Psi_G} \ket{\psi_a} \;\Rightarrow\; 
\overbrace{\ket{\Psi_G \rightarrow \omega}}^{\substack{n\ \text{qubits,}\\ \text{read}}} 
\underbrace{\ket{0}}_{\substack{1\ \text{ancilla}}} 
\;+\;
\overbrace{\ket{\Psi_G \rightarrow \emptyset}}^{\substack{n\ \text{qubits,}\\ \text{do not read}}} 
\underbrace{\ket{1}}_{\substack{1\ \text{ancilla}}},
\end{equation}
therefore, one reads out the result of all $n$ working qubits only if the ancilla is found in the $\ket{0}_{ancilla}$ state. Such an operation requires the following number of qubit readouts on average: 

\begin{equation}
    R^{\mathcal{PT}}\left(k\right)= \frac{1}{\mathcal{D}_G^{\mathcal{PT}}\left(k,\alpha\rightarrow\frac{\pi}{2}\right)} + n = \frac{1}{p_+\left(k\right) } + n,
\end{equation}
where $1$ corresponds to reading a single ancilla possibly several times until postselection is successful, and $n$ corresponds to the readout of the rest of the qubits when postselection succeeds. Such an approach uses a smaller number of qubit readouts compared to the conventional punctuated Grover's search algorithm:
\begin{equation}
    \frac{R^{\mathcal{PT}}\left(k\right)}{R\left(k\right)} = \frac{n \cdot p_+\left(k\right)+1}{n}\underbrace{\longrightarrow}_{n \gg 1} p_+\left(k\right) \le 1
\end{equation}
For example, when the punctuated quantum database search minimizing the number of oracle calls described in Eqn.~\eqref{T(k)} is run, our approach reduces the number of qubit readouts from $ \frac{n}{\sin^2\left(2\cdot 0.583\right)}= 1.184 \cdot n$ to $n$, and from $2n$ to $n$ when one is allowed to use half the depth of the original Grover's search cycle, in the $n \gg 1$ limit.

This advantage becomes even more significant and can outperform the conventional punctuated search in terms of qubit readouts by several times if one is allowed to use less than half the depth to complete the full cycle of Grover's search, as illustrated in Fig.~\ref{GrFig}. Additionally, in a similar manner, the number of qubit readouts can be reduced if several oracles are run in parallel to find one of several possible solutions~\cite{gingrich2000generalized}. Finally, our $\mathcal{PT}$-symmetric approach makes the depth-measurement trade-off discussed in~\cite{ng2024} redundant and unnecessary since all $n$ working qubits are measured \underline{once}, only after a successful post-selection.

\section{Conclusions and future work}
\label{Conclusions}

The main result of our paper is the identification of specific scenarios and performance metrics where $\mathcal{PT}$-symmetric systems outperform their Hermitian counterparts. This achievement was made possible by developing a new $\mathcal{PT}$-symmetric algorithm for mapping three arbitrary quantum states, deriving exact expressions for the probability of successful postselection, and verifying our theoretical computations using IBM Quantum Experience.

We demonstrated that, when applied to the discrimination of $N = 2$ states, our approach provides equivalent results to conventional unambiguous quantum state discrimination. When applied to an attack on the three-state QKD protocol, our approach yields the same error rate as other approaches in the literature, thus aligning with the security proof of this QKD protocol. However, our methodology can be advantageous in practical scenarios involving the discrimination of highly asymmetric quantum states.

Through explicit computations, we identified the conditions under which $\mathcal{PT}$-symmetric sensors outperform their Hermitian counterparts, thereby relaxing the conditions necessary to achieve this performance. Furthermore, we explicitly demonstrated that our $\mathcal{PT}$-symmetric approach surpasses the conventional punctuated quantum database search in terms of qubit readout cost, an important practical parameter, while maintaining the same average number of oracle calls. 

With our approach already implemented in an optical scheme~\cite{chen2022quantum}, our work lays the foundation for leveraging the unique properties of $\mathcal{PT}$ symmetry to advance quantum information processing, communication, cryptography, and sensing.

\section*{Acknowledgements}

This publication is based on work supported by a grant from the U.S. Civilian Research \& Development Foundation (CRDF Global), award number
G-202105-67836. YB was funded in part by the UCCS BioFrontiers Center and the UCCS Cyber Seed Grant. The authors express their gratitude for valuable discussions with Manohar Raavi and Anatoliy Pinchuk during the initial stages of this work.

\section*{Data and code availability}
The code, implementation, and results of the runs on the IBM Quantum Experience are publicly available at GitHub repository:

\href{https://github.com/BalytskyiJaroslaw/QuantumSimulations/tree/master}{https://github.com/BalytskyiJaroslaw/QuantumSimulations/tree/master}. 

\section{Methods}\label{Methods}

\subsection{Adjusting to convenient positions for \textit{Step 1}}
A set of arbitrary three states $|\psi_i\rangle = \begin{pmatrix}
\cos\left(\frac{\theta_i}{2}\right)  \\
e^{i\phi_i}\sin\left(\frac{\theta_i}{2}\right)
\end{pmatrix}, i \in \left[1, 3\right]$ can be  adjusted to the starting positions in Eqn.~\eqref{StartingPosition} by the following unitary rotation:
\begin{dmath}
\mathcal{R} =
\begin{pmatrix}
\cos\left(\frac{\pi - 2\sigma}{4}\right) & -i \sin\left(\frac{\pi - 2\sigma}{4}\right) \\
-i \sin\left(\frac{\pi - 2\sigma}{4}\right) & \cos\left(\frac{\pi - 2\sigma}{4}\right)
\end{pmatrix}
\cdot
\begin{pmatrix}
1 & 0 \\
0 & -i e^{-i\lambda - i\phi_2}
\end{pmatrix}
\cdot
\begin{pmatrix}
\cos\left(\frac{\theta_1}{2}\right) & \sin\left(\frac{\theta_1}{2}\right) e^{-i\phi_1} \\
-\sin\left(\frac{\theta_1}{2}\right) e^{i\phi_1} & \cos\left(\frac{\theta_1}{2}\right)
\end{pmatrix}
\label{Start}
\end{dmath}
The starting position parameters in Eqn.~\eqref{StartingPosition} are expressed as:
\begin{dmath}
\cos\left(\frac{\mu}{2}\right) = \lvert\beta\lvert  = \sqrt{\left(Re\left(\beta\right)\right)^2 + \left(Im\left(\beta\right)\right)^2},
\end{dmath}
\begin{dmath}
\nu = \arctan\left(\frac{Im\left(\gamma\right)}{Re\left(\gamma\right)}\right) - \arctan\left(\frac{Im\left(\beta\right)}{Re\left(\beta\right)}\right),
\end{dmath}
with $\sigma$, and $\lambda$ parameters given by the Eqns.~\eqref{PrepareHamiltonian1},~\eqref{PrepareHamiltonian2},~\eqref{PrepareHamiltonian3}, and~\eqref{PrepareHamiltonian4} as:
\begin{dmath}
\begin{aligned}\label{PrepareHamiltonian1}
\cos\left(\sigma\right) = \sqrt{\frac{1 +\cos(\text{$\theta_1$}) \cos (\text{$\theta_2$}) + \sin (\text{$\theta_1$})\sin (\text{$\theta_2$}) \cos(\text{$\phi_1$}-\text{$\phi_2$})}{2}}
  \end{aligned},
\end{dmath}
\begin{dmath}
  \begin{aligned}
\label{PrepareHamiltonian2}\lambda  =\arctan\left(\frac{\sin\left(\frac{\theta_1}{2}\right)\cos\left(\frac{\theta_2}{2}\right)\sin\left(\phi_2 - \phi_1\right)}{\cos\left(\frac{\theta_1}{2}\right)\sin\left(\frac{\theta_2}{2}\right) - \sin\left(\frac{\theta_1}{2}\right)\cos\left(\frac{\theta_2}{2}\right)\cos\left(\phi_2 - \phi_1\right)}\right) - \\- \arctan\left(\frac{\sin\left(\frac{\theta_1}{2}\right)\sin\left(\frac{\theta_2}{2}\right)\sin\left(\phi_2 - \phi_1\right)}{\cos\left(\frac{\theta_1}{2}\right)\cos\left(\frac{\theta_2}{2}\right) + \sin\left(\frac{\theta_1}{2}\right)\sin\left(\frac{\theta_2}{2}\right)\cos\left(\phi_2 - \phi_1\right)}\right),  \end{aligned}
\end{dmath}
\begin{dmath}
\begin{aligned}
\label{PrepareHamiltonian3}
\beta = \cos\left(\frac{\theta_1}{2}\right)\cos\left(\frac{\theta_3}{2}\right)\cos\left(\frac{\pi - 2\sigma}{4}\right)\left(1 + \tan\left(\frac{\theta_1}{2}\right)\tan\left(\frac{\pi - 2\sigma}{4}\right)e^{i\phi_1 - i\phi_2 - i\lambda}\right) + \\
+ \sin\left(\frac{\theta_1}{2}\right)\sin\left(\frac{\theta_3}{2}\right)\cos\left(\frac{\pi - 2\sigma}{4}\right)e^{i\phi_3 - i\phi_1}\left(1 - \cot\left(\frac{\theta_1}{2}\right)\tan\left(\frac{\pi - 2\sigma}{4}\right)e^{i\phi_1 - i\phi_2 - i\lambda}\right),
\end{aligned}
\end{dmath}
\begin{dmath}\label{PrepareHamiltonian4}
\begin{aligned}
\gamma = i\cos\left(\frac{\theta_1}{2}\right)\cos\left(\frac{\theta_3}{2}\right)\sin\left(\frac{\pi - 2\sigma}{4}\right)\left(\tan\left(\frac{\theta_1}{2}\right)\cot\left(\frac{\pi - 2\sigma}{4}\right)e^{i\phi_1 - i\phi_2 - i\lambda} - 1\right) - \\
- i\sin\left(\frac{\theta_1}{2}\right)\sin\left(\frac{\theta_3}{2}\right)\sin\left(\frac{\pi - 2\sigma}{4}\right)e^{i\phi_3 - i\phi_1}\left(1 + \cot\left(\frac{\theta_1}{2}\right)\cot\left(\frac{\pi - 2\sigma}{4}\right)e^{i\phi_1 - i\phi_2 - i\lambda}\right)
\end{aligned}
\end{dmath}

\subsection{Unitary rotation, \textit{Step 2}}
The unitary rotation, which prepares the states for \textit{Step 3} and subsequent $\mathcal{PT}$-symmetric evolution, is defined by the following parameters:
\begin{dmath}
\kappa = \cos\left(\frac{\mu}{2}\right)\left( \cos\left(\omega\tau - \alpha\right)\cos\left(\frac{\delta}{2}\right) + \sin\left(\omega\tau \right)\sin\left(\frac{\delta}{2}\right) \right) + i e^{i\nu} \sin\left(\frac{\mu}{2}\right)\left( \cos\left(\omega\tau + \alpha\right)\sin\left(\frac{\delta}{2}\right) - \sin\left(\omega\tau \right)\cos\left(\frac{\delta}{2}\right) \right),
\end{dmath}
\begin{dmath}
\zeta = i\cos\left(\frac{\mu}{2}\right)\left( \cos\left(\omega\tau - \alpha\right)\sin\left(\frac{\delta}{2}\right) - \sin\left(\omega\tau \right)\cos\left(\frac{\delta}{2}\right) \right) + e^{i\nu} \sin\left(\frac{\mu}{2}\right)\left( \cos\left(\omega\tau + \alpha\right)\cos\left(\frac{\delta}{2}\right) + \sin\left(\omega\tau \right)\sin\left(\frac{\delta}{2}\right) \right), 
\end{dmath}
\begin{equation}\label{Step2}
\cos\left(\frac{\xi}{2}\right)  = \frac{|\kappa|}{\sqrt{|\kappa|^2 + |\zeta|^2}}, \ \chi = \arctan\left(\frac{Im\left(\zeta\right)}{Re\left(\zeta\right)}\right) - \arctan\left(\frac{Im\left(\kappa\right)}{Re\left(\kappa\right)}\right)
\end{equation}

\subsection{Implementation of $\mathcal{PT}$ symmetry by the dilation method}\label{Implementation}

The combined qubit-ancilla system is governed by the following \textit{Hermitian} Hamiltonian:
\begin{equation}
H^{Total}_{a,q}\left(t\right) = \hat{1}\otimes \Sigma \left(t\right) + \sigma_y\otimes\Upsilon\left(t\right)\label{TotalH},
\end{equation}
and its elements are given by:
\begin{equation}
\Sigma\left(t\right) = \left[\mathcal{H}_q\left(t\right) + i\frac{\mathrm{d}\zeta\left(t\right)}{\mathrm{d}t}\zeta\left(t\right) + \zeta\left(t\right) \mathcal{H}_q\left(t\right)\zeta\left(t\right)\right]\mathcal{N}^{-1}\left(t\right),
\end{equation}
\begin{equation}
\Upsilon\left(t\right) = i\left[\mathcal{H}_q\left(t\right)\zeta\left(t\right) -\zeta\left(t\right)\mathcal{H}_q\left(t\right) - i\frac{\mathrm{d}\zeta\left(t\right)}{\mathrm{d}t} \right]\mathcal{N}^{-1}\left(t\right) ,
\end{equation}
\begin{equation}
\mathcal{N}\left(t\right) = T \exp\left[-i\int_0^t\mathrm{d}\tau \ \mathcal{H}^\dagger_q\left(\tau\right)\right]\mathcal{N}\left(0\right)\tilde{T}\exp\left[i\int^t_0 \mathrm{d}\tau \ \mathcal{H}\left(\tau\right)\right],
\end{equation}
where $T$ and $\tilde{T}$ are the time and and anti-time-ordering operators, respectively. The operator $\zeta\left(t\right) = \left(\mathcal{N}\left(t\right) - \hat{1}\right)^{\frac{1}{2}}$ must maintain all its eigenvalues real, and the initial value $\mathcal{N}\left(0\right)$ must be chosen accordingly to ensure it. The following system of equations:
\begin{equation}
\begin{cases}
\Sigma\left(t\right) - i\Upsilon\left(t\right)\zeta\left(t\right)= \mathcal{H}_q\left(t\right) \\
\Sigma\left(t\right)\zeta\left(t\right) + i\Upsilon\left(t\right) = i\frac{\mathrm{d}\zeta\left(t\right)}{\mathrm{d}t} + \zeta\left(t\right)\mathcal{H}_q\left(t\right)
\end{cases},
\end{equation}
ensures that the driven qubit is evolved by the $\mathcal{PT}$-symmetric Hamiltonian in Eqn.~\eqref{Hamiltonian1}. The ancilla qubit must be initialized as $|\psi\left(0\right)\rangle_{a} = \frac{1}{\sqrt{\zeta\left(0\right)^2 + 1}}\left(|0\rangle_a + \zeta\left(0\right)|1\rangle_a\right)$. For both stages of the $\mathcal{PT}$-symmetric evolution, $4 \times 4$ evolution matrix was obtained by numerical solution of differential equations by Mathematica~\cite{Mathematica}. Finally, the evolution matrices $U_{Evolution}$ were decomposed into the elementary gates U3 as defined by IBM. For the first part of the $\mathcal{PT}$-symmetric evolution, they are denoted as $U^i_j$, and $V^i_j$ for the second part, where $i,j\in\left[1,4\right]$. This was done employing the method defined in~\cite{vatan2004optimal,sousa2006universal}. First, the rotation to the ``magic basis'' defined as:

\begin{equation}
\begin{cases}
\lvert\phi_1\rangle = \frac{1}{\sqrt{2}}\left(\lvert 00\rangle + \lvert 11\rangle \right); \ \ \lvert\phi_2\rangle = \frac{-i}{\sqrt{2}}\left(\lvert 00\rangle - \lvert 11\rangle \right) \\
\lvert\phi_3\rangle = \frac{1}{\sqrt{2}}\left(\lvert 01\rangle - \lvert 10\rangle \right) ; \ \
\lvert\phi_4\rangle = \frac{-i}{\sqrt{2}}\left(\lvert 01\rangle + \lvert 10\rangle \right)
\end{cases},
\end{equation}
was performed. As a result, the evolution matrix was factorized as:

\begin{equation}
U_{Evolution} = \left(U_A\otimes U_B\right)\cdot U_D\cdot \left(V_A\otimes V_B\right),
\end{equation}
\begin{equation}
U_D = e^{i\theta_0}MatrixExp\left(i\sum_{k = 1}^3 \theta_k\sigma_k\otimes\sigma_k\right) = \sum_{k = 1}^4 e^{I\Phi_k}\lvert\phi_k\rangle\langle\phi_k\lvert
\end{equation}
The final transformation is carried out by $M$ and $\Lambda$ matrices:

\begin{equation}
M = \frac{1}{\sqrt{2}}\begin{pmatrix}
 1 & 0 & 0 & i \\
 0 & i & 1 & 1 \\
 0 & i & -1 & 0 \\
 1 & 0 & 0 & -i
\end{pmatrix}, \
\Lambda = \begin{pmatrix}
 1 & 1 & -1 & 1 \\
 1 & 1 & 1 & -1 \\
 1 & -1 & -1 & -1 \\
 1 & -1 & 1 & 1
\end{pmatrix},
\end{equation}
\begin{equation}
\theta = \left(\theta_0, \theta_1, \theta_2, \theta_3\right)^T; \ \Phi = \left(\Phi_0, \Phi_1, \Phi_2, \Phi_3\right)^T; \ \theta = \Lambda\cdot\Phi
\end{equation}

The numerical results are as follows.

\subsubsection{First $\mathcal{PT}$-symmetric evolution for $\sigma = \frac{4}{5}$, $\alpha = \frac{\pi}{2} - 1$, $\mathcal{N}\left(0\right) =  3$}

\begin{figure}[t]
\centering
\scalebox{0.9}{
\begin{quantikz}
& \lstick{$\ket{0}$} & \gate{\mathcal{A}_1} & \gate{U^2_1} \gategroup[2,steps=7,style={dashed,
rounded corners,fill=blue!20, inner xsep=2pt},
background]{{\sc $\mathcal{PT}$-symmetric evolution \#1}}&\ctrl{1} & \gate{U^2_2} & \ctrl{1} & \gate{U^2_3} & \ctrl{1} & \gate{U^2_4}  & \meter[draw=red]{Postselection $\ket{0}$} & \qw  \\
& \lstick{$\ket{\psi_{\left(1,2,3\right)}}$} & \qw & \gate{U_1^1} & \targ{} & \gate{U_2^1} & \targ{} &  \gate{U^1_3} & \targ{} & \gate{U_4^1} & \rstick{$\ket{\psi_{\left(1,2,3\right)}}$ evolved to \\ $\cos^2\left(\kappa_{12}\right) = 0$}\qw 
\end{quantikz}
}
\caption{\ \ \ \textit{Step 1} and the first stage of the $\mathcal{PT}$-symmetric evolution.}
\label{Part1Evolve}
\end{figure}
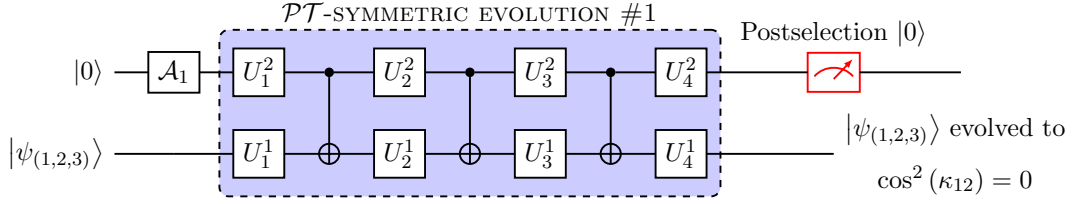

$U^{First Stage} =$ 
\begin{center}
\resizebox{\textwidth}{!}{
$
\begin{bmatrix}
 0.142552\, -0.235663 i & -0.650522-0.393504 i & 0.2897\, -0.478919 i & -0.155267-0.0939248 i \\
 -0.650482-0.393478 i & 0.257979\, -0.426478 i & -0.155257-0.0939158 i & -0.194071+0.320825 i \\
 -0.289696+0.478909 i & 0.155284\, +0.0939324 i & 0.142546\, -0.23565 i & -0.650547-0.393515 i \\
 0.15525\, +0.0939129 i & 0.194076\, -0.320837 i & -0.650472-0.393474 i & 0.257971\, -0.42646 i \\
\end{bmatrix}
$
}
\end{center}

\begin{equation}
\left(\Phi_0,\Phi_1,\Phi_2,\Phi_3\right)  = \left(1.61364,1.61364 ,2.61598,2.61598\right),
\end{equation}

\begin{equation}
U_A = \left(
\begin{array}{cc}
 0.553173 -0.0868701 i & 0.128532 +0.818494 i \\
 0.818494 -0.128536 i & -0.0868671-0.553173 i \\
\end{array}
\right),
\end{equation}

\begin{equation}
U_B = \left(
\begin{array}{cc}
 0.27538\, -0.495893 i & 0.823563 \\
 0.252434\, -0.783922 i & -0.556431+0.110127 i \\
\end{array}
\right),
\end{equation}

\begin{equation}
V_A = \left(
\begin{array}{cc}
 -0.35228 i & -0.9359 i \\
 -0.9359 & 0.35228 \\
\end{array}
\right),
\end{equation}

\begin{equation}
V_B = \left(
\begin{array}{cc}
 0.1277\, -0.81357 i & 0.19514\, -0.53266 i \\
 0.53266\, -0.19514 i & -0.81357+0.1277 i \\
\end{array}
\right)
\end{equation}

\subsubsection{First $\mathcal{PT}$-symmetric Evolution for $\sigma = \frac{6}{5}$, $\alpha = \frac{\pi}{2} - 1$, $\mathcal{N}\left(0\right) =  2$}

$U^{First Stage} =$ 
\begin{center}
\resizebox{\textwidth}{!}{
$
\begin{bmatrix}
 0.49542\, -0.32066 i & -0.28774-0.44456 i & 0.50219\, -0.32504 i & -0.06318-0.09762 i \\
 -0.28768-0.44447 i & 0.66023\, -0.42733 i & -0.06317-0.0976 i & -0.24838+0.16076 i \\
 -0.50223+0.32506 i & 0.06318\, +0.09761 i & 0.49546\, -0.32069 i & -0.28773-0.44455 i \\
 0.06318\, +0.09762 i & 0.24838\, -0.16076 i & -0.28769-0.44449 i & 0.66024\, -0.42733 i \\
\end{bmatrix}
$
}
\end{center}

\begin{equation}
\left(\Phi_0,\Phi_1,\Phi_2,\Phi_3\right)  = \left(-0.09904, -0.09904, -1.04982, -1.04984\right),
\end{equation}

\begin{equation}
U_A = \left(
\begin{array}{cc}
 0.32195\, +0.35488 i & -0.65003+0.5898 i \\
 0.58977\, +0.65006 i & 0.35485\, -0.32198 i \\
\end{array}
\right),
\end{equation}

\begin{equation}
U_B = \left(
\begin{array}{cc}
 -0.68589+0.4424 i & 0.57779 \\
 -0.57506+0.05604 i & -0.72556-0.37379 i \\
\end{array}
\right),
\end{equation}

\begin{equation}
V_A = \left(
\begin{array}{cc}
 0.\, -0.27972 i & -0.00004-0.96008 i \\
 -0.96008-0.00002 i & 0.27972 \\
\end{array}
\right),
\end{equation}

\begin{equation}
V_B = \left(
\begin{array}{cc}
 -0.42795-0.38828 i & -0.78849-0.21068 i \\
 0.21067\, +0.78849 i & -0.38827-0.42795 i \\
\end{array}
\right)
\end{equation}

\subsubsection{Second $\mathcal{PT}$-symmetric Evolution, $\alpha = \frac{\pi}{2} - 1$}
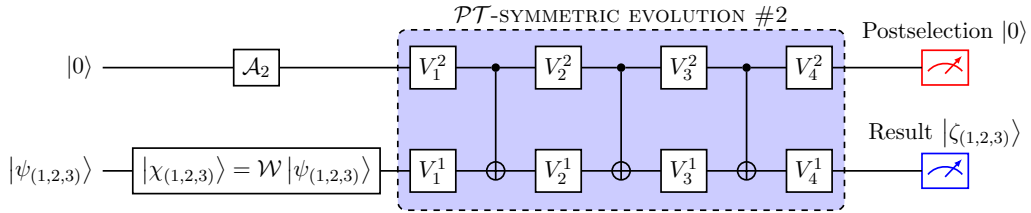
\begin{figure}[ht]
\centering
\scalebox{0.8}{
\begin{quantikz}
& \lstick{$\ket{0}$} & \gate{\mathcal{A}_2} & \gate{V^2_1} \gategroup[2,steps=7,style={dashed,
rounded corners,fill=blue!20, inner xsep=2pt},
background]{{\sc $\mathcal{PT}$-symmetric evolution \#2}}&\ctrl{1} & \gate{V^2_2} & \ctrl{1} & \gate{V^2_3} & \ctrl{1} & \gate{V^2_4}  & \meter[draw=red]{Postselection $\ket{0}$}  \\
& \lstick{$\ket{\psi_{\left(1,2,3\right)}}$} & \gate{\ket{\chi_{\left(1,2,3\right)}}=\mathcal{W}\ket{\psi_{\left(1,2,3\right)}}} & \gate{V_1^1} & \targ{} & \gate{V_2^1} & \targ{} &  \gate{V^1_3} & \targ{} & \gate{V_4^1} &  \meter[draw=blue]{Result $\ket{\zeta_{\left(1,2,3\right)}}$}
\end{quantikz}
}
\caption{\ \ \ \textit{Steps 2} and \textit{3}. Unitary rotation $\mathcal{W}$ puts the evolved state into conventional positions, and the second $\mathcal{PT}$-symmetric evolution eliminates one of three states or reduces them to the mirror-symmetric ones.}
\label{Part2Evolve}
\end{figure}

\begin{equation}
U^{Second \ stage}=\left(
\begin{array}{cccc}
 0.20775 i & -0.75471 & 0.55429 i & 0.28286 \\
 -0.75471 & -0.2077 i & 0.28292 & -0.55429 i \\
 -0.55429 i & -0.28286 & 0.20775 i & -0.75471 \\
 -0.28292 & 0.55429 i & -0.75471 & -0.2077 i \\
\end{array}
\right)
\end{equation}

\begin{equation}
U_A = \left(
\begin{array}{cc}
 -0.35523 - 0.0003 i & -0.94791 - 0.00081 i \\
 0.94791 + 0.00081 i & -0.35523 - 0.0003 i \\
\end{array}
\right),
\end{equation}

\begin{equation}
U_B = \left(
\begin{array}{cc}
 0.00085 - i & 0.00005 \\
 -0.00005 & -0.00085 + i \\
\end{array}
\right),
\end{equation}

\begin{equation}
V_A = V_B = \hat{1}; \left(\Phi_0,\Phi_1,\Phi_2,\Phi_3\right) = \left(0.93734, -0.93734, 0.93734, -0.93734\right)
\end{equation}

\subsubsection{Second $\mathcal{PT}$-symmetric Evolution, $\alpha = \frac{\pi}{2} - 0.7$}

\begin{equation}
U^{Second \ stage}=\left(
\begin{array}{cccc}
 0.29118 i & -0.68017 & 0.57944 i & 0.3418 \\
 -0.68017 & -0.29097 i & 0.34196 & -0.57945 i \\
 -0.57944 i & -0.3418 & 0.29118 i & -0.68017 \\
 -0.34197 & 0.57946 i & -0.68016 & -0.29098 i \\
\end{array}
\right)
\end{equation}

\begin{equation}
U_A = \left(
\begin{array}{cc}
 -0.44888 - 0.00094 i & -0.89359 - 0.00187 i \\
 0.89359 + 0.00187 i & -0.44888 - 0.00094 i \\
\end{array}
\right),
\end{equation}

\begin{equation}
U_B = \left(
\begin{array}{cc}
 -0.0021 - i & 0.00011 \\
 -0.00011 & 0.0021 + i \\
\end{array}
\right),
\end{equation}

\begin{equation}
V_A = V_B = \hat{1}; \left(\Phi_0,\Phi_1,\Phi_2,\Phi_3\right) = \left(0.86525, -0.86525, 0.86525, -0.86525\right)
\end{equation}

\subsubsection{Second $\mathcal{PT}$-symmetric Evolution, $\alpha = \frac{\pi}{2} - 0.5$}

\begin{equation}
U^{Second \ stage}=\left(
\begin{array}{cccc}
 0.3604 i & -0.62104 & 0.5786 i & 0.38684 \\
 -0.6211 & -0.36067 i & 0.3866 & -0.57852 i \\
 -0.5786 i & -0.38684 & 0.36039 i & -0.62105 \\
 -0.3866 & 0.57852 i & -0.62111 &  -0.36067 i \\
\end{array}
\right)
\end{equation}

\begin{equation}
U_A = \left(
\begin{array}{cc}
 0.52887 & 0.8487 +0.00001 i \\
 -0.8487-0.00001 i & 0.52887 \\
\end{array}
\right),
\end{equation}

\begin{equation}
U_B = \left(
\begin{array}{cc}
 0.00001 + i & 0.00017 \\
 -0.00017 & -0.00001 - i \\
\end{array}
\right),
\end{equation}

\begin{equation}
V_A = V_B = \hat{1}; \left(\Phi_0,\Phi_1,\Phi_2,\Phi_3\right) = \left(0.82071, -0.82071, 0.82071, -0.82072\right)
\end{equation}

The first part of the $\mathcal{PT}$-symmetric evolution is shown in Fig.~\ref{Part1Evolve}, and the unitary rotation with the second part of the $\mathcal{PT}$-symmetric evolution in Fig.~\ref{Part2Evolve} respectively. 

\subsubsection{Attack on the three-state QKD protocol}

\begin{equation}
U^{Three \ State \ QKD}=\left(
\begin{array}{cccc}
 -0.09739 i & -0.87214 & 0.43866 i & -0.1937 \\
 -0.87215 & 0.0974 i & -0.19372 & -0.43864 i \\
 -0.4386 i & 0.19367 & -0.09744 i & -0.87212 \\
 0.19366 & 0.43861 i & -0.8721 & 0.09742 i \\
\end{array}
\right)
\end{equation}

\begin{equation}
U_A = \left(
\begin{array}{cc}
 -0.2167 + 0.00644 i & 0.97579 - 0.02899 i \\
 -0.97579 + 0.02899 i & -0.2167 + 0.00644 i \\
\end{array}
\right),
\end{equation}

\begin{equation}
U_B = \left(
\begin{array}{cc}
 -0.02969 + 0.99956 i & 0.00001 \\
 -0.00001 & 0.02969 - 0.99956 i \\
\end{array}
\right),
\end{equation}

\begin{equation}
V_A = V_B = \hat{1}; \left(\Phi_0,\Phi_1,\Phi_2,\Phi_3\right) = \left(1.10479, -1.10479, 1.10482, -1.10482 \right)
\end{equation}

\bibliographystyle{plainnat}

\end{document}